\documentclass{article}

\usepackage{arxiv}

\usepackage{amssymb}

\usepackage{graphicx}
\usepackage[space]{grffile}
\usepackage{latexsym}
\usepackage{textcomp}
\usepackage{longtable}
\usepackage{tabulary}
\usepackage{booktabs,array,multirow}
\usepackage{amsfonts,amsmath,amssymb}
\usepackage{url}
\usepackage{hyperref}
\hypersetup{colorlinks=true,linkcolor=red,citecolor=black,urlcolor=magenta}
\usepackage{etoolbox}
\usepackage{natbib}
\usepackage{setspace}
\usepackage{placeins}
\usepackage[title]{appendix}

\usepackage{lineno} 

\newif\iflatexml\latexmlfalse

\AtBeginDocument{\DeclareGraphicsExtensions{.pdf,.PDF,.eps,.EPS,.png,.PNG,.tif,.TIF,.jpg,.JPG,.jpeg,.JPEG}}

\usepackage[utf8]{inputenc}
\usepackage[english]{babel}
\usepackage{siunitx}
\usepackage[T1]{fontenc} 
\usepackage{bm}
\usepackage{mathtools}
\usepackage{nicefrac} 
\usepackage{microtype}
\usepackage{lipsum}
\usepackage[ruled,vlined]{algorithm2e}
\usepackage{xcolor}
\usepackage{afterpage}
\usepackage{pdflscape}
\usepackage[normalem]{ulem}

\usepackage[textsize=tiny]{todonotes}
\definecolor{org}{rgb}{1,0.53,0.0}

\definecolor{mrg}{rgb}{0.1,0.45,0.1}

\usepackage{amsthm}
\newtheorem{remark}{Remark}
\newtheorem{proposition}{Proposition}

\newcommand{\w}{\bm{w}}
\newcommand{\x}{\bm{x}}
\newcommand{\y}{\bm{y}}
\newcommand{\z}{\bm{z}}

\newcommand{\X}{\mathbf{X}}
\newcommand{\Y}{\mathbf{Y}}
\newcommand{\Z}{\mathbf{Z}}
\newcommand{\W}{\mathbf{W}}

\newcommand{\A}{\mathbf{A}}
\newcommand{\B}{\mathbf{B}}
\newcommand{\C}{\mathbf{C}}

\newcommand{\K}{\mathbf{K}}
\newcommand{\bfc}{\mathbf{c}}
\newcommand{\0}{\mathbf{0}}

\newcommand{\bfSigma}{\boldsymbol{\Sigma}}

\newcommand{\T}{\bm{T}}
\renewcommand{\S}{\bm{S}}

\renewcommand{\a}{\bm{a}}
\renewcommand{\b}{\bm{b}}
\newcommand{\SKR}{\bm{S}}

\newcommand{\KLDiv}{\mathcal{D}_{\textrm{KL}}}
\newcommand\ci{\perp\!\!\!\perp}
\newcommand{\R}{\mathbb{R}}

\title{Ensemble transport smoothing\\ Part I: Unified framework}
\renewcommand{\shorttitle}{Ensemble transport smoothing, Part I}

\author{Maximilian Ramgraber\thanks{Authors contributed equally to this work} \\
	Department of Aeronautics and Astronautics\\
	Massachusetts Institute of Technology\\
	Cambridge, MA 02139 \\
	\texttt{mramgrab@mit.edu} \\
	\And
	Ricardo Baptista\footnotemark[1] \\
	Department of Aeronautics and Astronautics\\
	Massachusetts Institute of Technology\\
	Cambridge, MA 02139 \\
	\texttt{rsb@mit.edu} \\
	\And
    Dennis McLaughlin  \\
    Department of Civil and Environmental Engineering\\
    Massachusetts Institute of Technology\\
    Cambridge, MA 02139 \\
    \texttt{dennism@mit.edu} \\
	\And
	Youssef Marzouk \\
	Department of Aeronautics and Astronautics\\
	Massachusetts Institute of Technology\\
	Cambridge, MA 02139 \\
	\texttt{ymarz@mit.edu} \\
}

\begin{document}

\maketitle

\begin{abstract}
Smoothers are algorithms for Bayesian time series re-analysis. Most operational smoothers rely either on affine Kalman-type transformations or on sequential importance sampling. These strategies occupy opposite ends of a spectrum that trades computational efficiency and scalability for statistical generality and consistency: non-Gaussianity renders affine Kalman updates inconsistent with the true Bayesian solution, while the ensemble size required for successful importance sampling can be prohibitive. This paper revisits the smoothing problem from the perspective of measure transport, which offers the prospect of consistent prior-to-posterior transformations for Bayesian inference. We leverage this capacity by proposing a general ensemble framework for transport-based smoothing. Within this framework, we derive a comprehensive set of smoothing recursions based on nonlinear transport maps and detail how they exploit the structure of state-space models in fully non-Gaussian settings. We also describe how many standard Kalman-type smoothing algorithms emerge as special cases of our framework. A companion paper \citep{Ramgraber2022underUpdatesb} explores the implementation of nonlinear ensemble transport smoothers in greater depth.
\end{abstract}

\keywords{Data assimilation \and smoothing \and ensemble methods \and triangular transport}

\section{Introduction}
Smoothing is a data assimilation technique focussed on retrieving improved estimates of a time series of states. Whereas \textit{filters} limit themselves to estimating only the current state of a dynamical system, smoothers also re-analyze past states, improving previous state estimates retroactively. This makes smoothers highly useful in systems where hindsight is also of interest. Both classes of algorithms find application in high-dimensional environmental systems \citep{Khare2008AnSystems}, tomographic imaging \citep{Hakkarainen2019UndersampledFilter}, and signal tracking \citep{Angelosante2009Lasso-KalmanSignals}.

In this paper, we revisit the smoothing problem from the perspective of measure transport \citep{Marzouk2017SamplingIntroduction,Spantini2022CouplingFiltering,Spantini2018InferenceCouplings,Villani2007OptimalNew}. Transport methods---discussed below---offer the prospect of \textit{consistent} prior-to-posterior transformations for Bayesian inference. We leverage this capacity by proposing a general ensemble framework for transport-based Bayesian smoothing. This transport framework unifies many previous approaches; in particular, many seemingly disparate Kalman-type smoothing algorithms emerge as linear special cases. Yet our perspective is quite general: we develop a comprehensive collection of nonlinear smoothing recursions and detail how they exploit the structure of state-space models in fully non-Gaussian settings. Adopting this unifying perspective, we also examine the strengths and drawbacks of different smoothing recursions. 

Smoothers currently comprise a wide and diverse range of algorithms, differing in update strategy (linear--Gaussian,\footnote{In this study, we will use the term \textit{linear--Gaussian} or \textit{Kalman-type} to refer to Bayesian updating (i.e., conditioning) operations that employ affine transformations derived from first and second moments of an ensemble. Such operations yield consistent solutions for linear dynamics and additive Gaussian noise distributions; otherwise, in general, they are approximations of the Bayesian solution.} variational, weight-based), order of operation (batch or sequential), and practical implementation (offline or online). The unifying element of these algorithms is their objective: characterizing the posterior distribution $p(\x_{1:t}|\y_{1:t}^{*})$ of sequences of model states $\x_{1:t} \coloneqq (\x_{1},\dots,\x_{t})$ given a time series of observations $\y_{1:t}^{*} \coloneqq (\y_{1}^{*},\dots,\y_{t}^{*})$. 
This posterior is obtained from the joint distribution $p(\x_{1:t},\y_{1:t})$ of a sequence of simulated model states $\x_{1:t}$ and their corresponding simulated observations $\y_{1:t}$ by conditioning on $\y_{1:t}^*$. The conditioning operation can be realized in bulk, as is common in 4D-variational inference \citep{LeDimet1986VariationalAspects} or the global variant of the ensemble smoother \citep{VanLeeuwen1996DataFormulation,Cosme2012SmoothingSolutions}.

\begin{figure}[!ht]
  \centering
  \includegraphics[width=\textwidth]{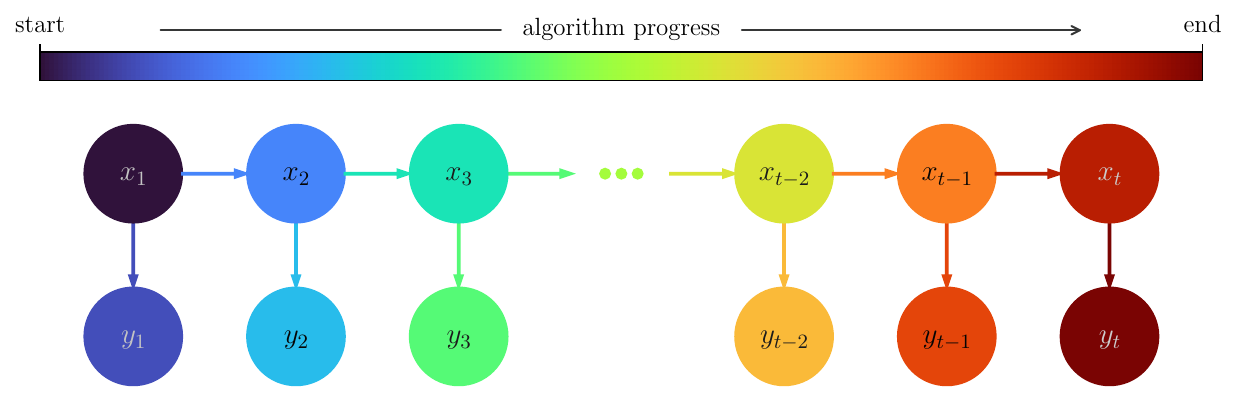}
  \caption{Graphical structure of a state-space or hidden Markov model (equivalent terms for our purposes). The joint distribution $p(\x_{1:t},\y_{1:t})$ unfolds sequentially via alternating applications of the forecast and observation models. $p(\x_{1})$ expands to $p(\x_{1},\y_{1})$, then $p(\x_{1:2},\y_{1})$, etc. The graphical model represents the conditional independence of the observations given the latent states, and the Markovian nature of the state sequence itself.}
  \label{fig:Markovian_graph}
\end{figure}

Most smoothers, however, operate sequentially instead. In dynamical settings, it is common to posit a prior distribution for the initial condition $p(\x_{1})$, a stochastic forecast model $p(\x_{s+1}|\x_{s})$, and a stochastic observation model $p(\y_{s}|\x_{s})$. Under the conditional independence assumptions encoded in a \textit{state-space} or \textit{hidden Markov} model (see Figure~\ref{fig:Markovian_graph}) \citep{ihler2007graphical,elliott2008hidden}, 
these ingredients are sufficient to specify the joint distribution 
$$p(\x_{1:t},\y_{1:t}) = p(\x_1) \prod_{s=1}^t p(\y_s \vert \x_s) \prod_{s=2}^t p(\x_{s} \vert \x_{s-1}),$$ 
which can be understood as unfolding sequentially in time by alternating applications of the forecast and observation models. Many filters and smoothers also sequence their conditioning operations accordingly, directly combining applications of the forecast model with incremental conditioning operations. The resulting recursion is commonly interpreted in terms of alternating \textit{forecast} and \textit{analysis} steps.

\begin{figure}[!ht]
  \centering
  \includegraphics[width=\textwidth]{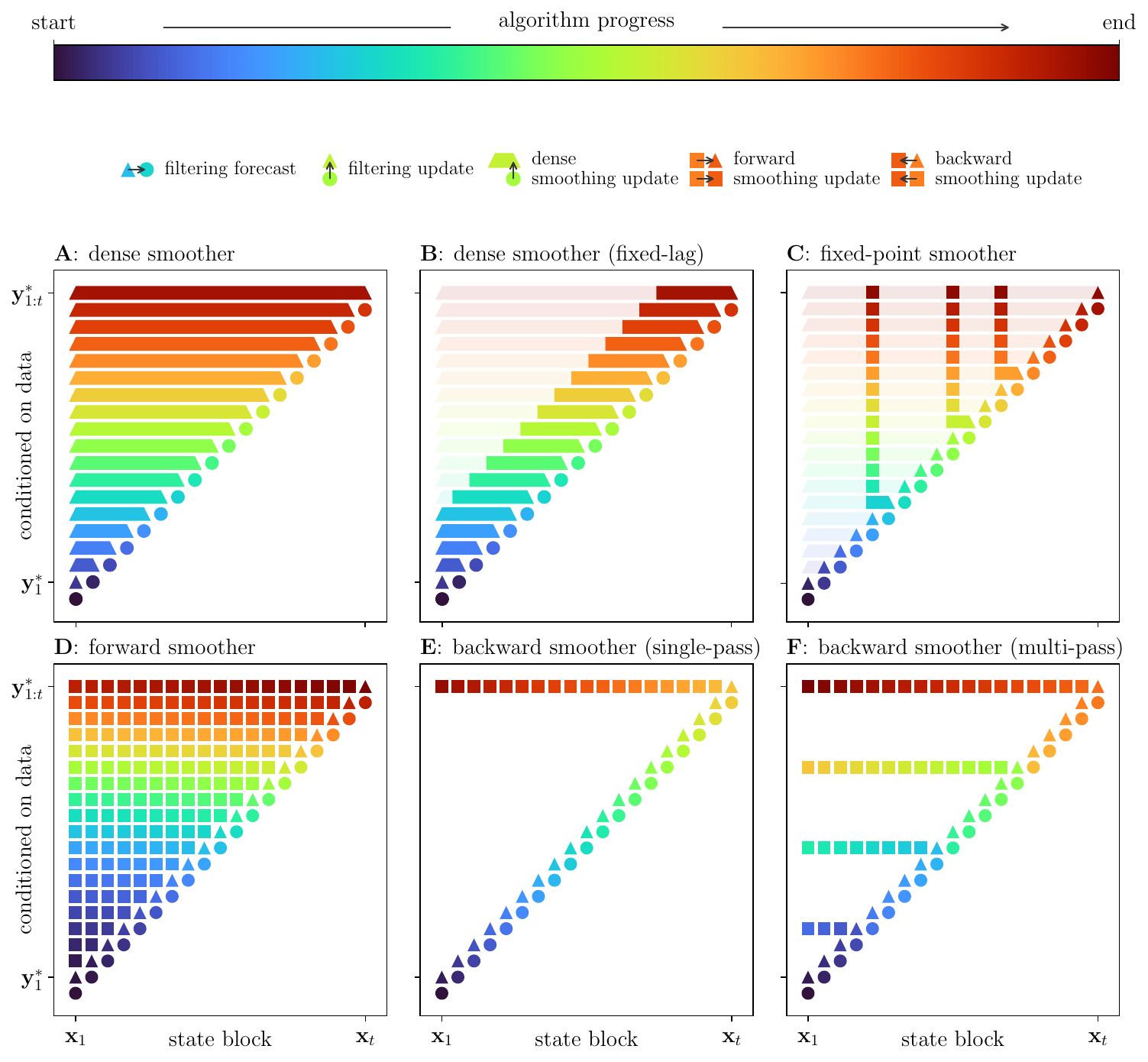}
  \caption{Joint and marginal distributions recovered by different smoothing algorithms. A block with coordinates $(\x_{i},\y_{1:j}^{*})$ represents samples from $p(\x_{i}|\y_{1:j}^{*})$. Color denotes the order of operations, and shape defines the operation by which each new block is obtained. A generic dense smoother (A) operates on a widening inference horizon, extending the analysis update to all past states. Its fixed-lag (B) and fixed-interval (C) variants only apply the update partially or selectively. Forward smoothers (D) update the states in chronological order at every timestep. Backward smoothers are based on filters but initiate backward passes either once at the end (single-pass, E) or intermittently (multi-pass, F).}
  \label{fig:smoother_types}
\end{figure}

The resulting recursive updates can be realized with different smoothing strategies. The simplest update strategy is a generic, \textit{dense} smoother (Figure~\ref{fig:smoother_types}A), which directly extract the conditional $p(\x_{1:t}|\y_{1:t-1}^{*}, \y_{t}^{*})$ from the joint distribution $p(\y_{t},\x_{1:t}|\y_{1:t-1}^{*})$ in a single inference operation without regard for any potential graphical structure. Through marginalization, it is possible to formulate \textit{fixed-lag} (Figure~\ref{fig:smoother_types}B) and \textit{fixed-point} (Figure~\ref{fig:smoother_types}C) variants of this smoother. In contrast, drawing on the state-space model's graphical structure allows us to formulate more advanced smoothing strategies: \textit{forward} (Figure~\ref{fig:smoother_types}D) or \textit{backward} smoothers (Figure~\ref{fig:smoother_types}E and F) exploit conditional independence relationships to realize serial updates, either forward-in-time or backward-in-time along the graph. We will discuss these methods in greater detail in Section~\ref{sec:ents_all}.

The practical realization of such Bayesian updates requires an implementation strategy. While closed-form solutions are available in certain restricted settings (e.g., conjugate prior/likelihood families), the need for broader applicability and computational tractability motivates the use of Monte Carlo approximations instead. Implementing such sample-based updates can be realized in different ways: \textit{sequential Monte Carlo} methods based on importance sampling and resampling \citep{Doucet2009ALater,Klaas2006FastParticles} use weighted ensembles and can reproduce arbitrarily complex nonlinear updates, but generally require vast ensemble sizes to forestall an inevitable weight collapse \citep{Snyder2008ObstaclesFiltering,Snyder2015PerformanceProposal}. Ensemble \textit{Kalman}-type methods \citep[e.g., ][]{Asch2016DataApplications,Evensen2003TheImplementation,Evensen2000AnDynamics} update prior to posterior through more sample-efficient ensemble transformations, but are limited to linear (more precisely, affine) updates, which are in general consistent only if all quantities involved are jointly Gaussian. In the context of smoothing, these techniques underlie algorithms such as the ensemble Kalman smoother \citep{Evensen2000AnDynamics}, and a broad range of algorithmic variants such as the ensemble smoother with multiple data assimilation \citep{Emerick2013EnsembleAssimilation} or the iterative ensemble Kalman smoother \citep{Bocquet2014AnSmoother,Evensen2019EfficientMatching}.

\textit{Transportation of measure} \citep[e.g.,][]{ElMoselhy2012BayesianMaps,Marzouk2017SamplingIntroduction,Pulido2019KernelFilter,tamang2021ensemble} 
provides a third option to realize ensemble smoothing. 
One approach involves using discrete optimal transport solvers to derive transformations of the ensemble that approximate the desired filtering or smoothing updates. Such approaches, e.g., as in \citet{reich2013nonparametric} and \citet{de2020ensemble}, require weighted ensembles, however, and hence are in the class of importance sampling methods that we do not consider here. Alternatively, recent work by \citet{Spantini2022CouplingFiltering} successfully leverages ensemble transport methods to establish a nonlinear generalization of the \textit{ensemble Kalman filter} (EnKF) \citep{Evensen1994SequentialStatistics,Burgers1998AnalysisFilter,houtekamer1998data}---an \textit{ensemble transport filter } (EnTF)---that is learned without weights. In this paper, we propose analogous nonlinear generalizations for ensemble smoothing. 

To this end, we first develop a unified transport-based framework for inference in state-space models, from which we derive a variety of nonlinear smoothing recursions---i.e., systematic ways of constructing transformations, all-at-once or sequentially, that realize the Bayesian update. These are collectively called ensemble transport smoothers (EnTS). We begin by revisiting the basics of ensemble transport in Sections~\ref{subsec:brief_intro}--\ref{sec:condsamp}; then, in Section~\ref{subsec:linear_maps}, we discuss how the affine Kalman-type update emerges as a special case of the more general ensemble transport update.

Within this framework, we subsequently elucidate differences between dense smoothers (Section~\ref{sec:any_ordering}), forward smoothers (Section~\ref{sec:ordering_A}), backward smoothers (Section~\ref{sec:ordering_B}), and fixed-point smoothers (Section~\ref{subsec:fixedpoint}), regarding both the flow and execution of the algorithms and the conditional independence properties that they exploit. We also show how canonical Kalman-type smoothers emerge as affine special cases of certain EnTS recursions, and identify important features (e.g., extra decomposability) that are unique to the linear--Gaussian case. 

Finally, we demonstrate these algorithmic equivalences numerically in Section~\ref{sec:numerics} and comment on how different smoothing recursions can differ in their finite-sample performance in Section~\ref{sec:conclusions}. A companion paper to this manuscript explores the practical implementation of nonlinear transport smoothing, along with localization and other regularization schemes, in greater depth \citep{Ramgraber2022underUpdatesb}. Notation used in this manuscript is summarized in Table~\ref{tab:nomenclature}.

\begin{table}
    \caption{Variables used in this study.}
    \begin{center}
    \begin{tabular}{ r l }
     $x$ or $S(x)$ & scalar-valued variables or functions \\ 
     $\x$ or $\SKR (\x)$ & vector-valued variables or functions \\  
     $\x$ & state variables \\
     $\y$ & predicted (unrealized) observations \\ 
     $\y^{*}$ & realized values of observations \\ 
     $N$ & ensemble size \\
     $\X$ or $\Y$ & ensemble representations of $\x$ or $\y$ (e.g., $\X = \{\x^i\}_{i=1}^N$) \\
     $\w$ or $\W$ & generic random variable and its ensemble representation \\
     $x \sim p$ & $x$ is distributed according to the probability distribution $p$ \\
     $1 \leq s \leq t$ & time step (often in a subscript) \\
     $1 \leq k \leq K$ & vector component index (often in a subscript) \\
     $1 \leq n \leq N$ & ensemble sample index (often in a superscript) \\
     $\X_{t}^{*}$ & $\X_{t}$ conditioned on $\y_{t}^{*}$; i.e., if $\X_{t}\sim p(\x_{t}|\y_{1:t-1}^{*})$, then $\X_{t}^{*}\sim p(\x_{t}|\y_{1:t}^{*})$ \\
     $p$ & target distribution\\
     $\eta$ & reference distribution; usually standard Gaussian $\mathcal{N}(\mathbf{0},\mathbf{I})$ \\
     $\SKR$ & transport map \\
     $\mathbf{C}$ & lower triangular matrix representing a linear transport map \\
     $\T$ & composite transport map\\
     $\SKR_{\sharp}p$ & pushforward distribution \\
     $\SKR^{\sharp}\eta$ & pullback distribution \\
    \end{tabular}
    \end{center}
    \label{tab:nomenclature}
\end{table}

\section{Transport, sparsity, and conditional sampling} \label{sec:transport_maps}
In the following, we will review the basics of measure transport methods, introducing the properties which permit conditional sampling and thus general Bayesian inference. We refer a reader interested in further mathematical detail to \citet{Marzouk2017SamplingIntroduction}, \citet{Spantini2018InferenceCouplings}, and \citet{ElMoselhy2012BayesianMaps}.

\subsection{Basic concepts of triangular transport}\label{subsec:brief_intro}

\begin{figure}[!ht]
  \centering
  \includegraphics[width=0.8\textwidth]{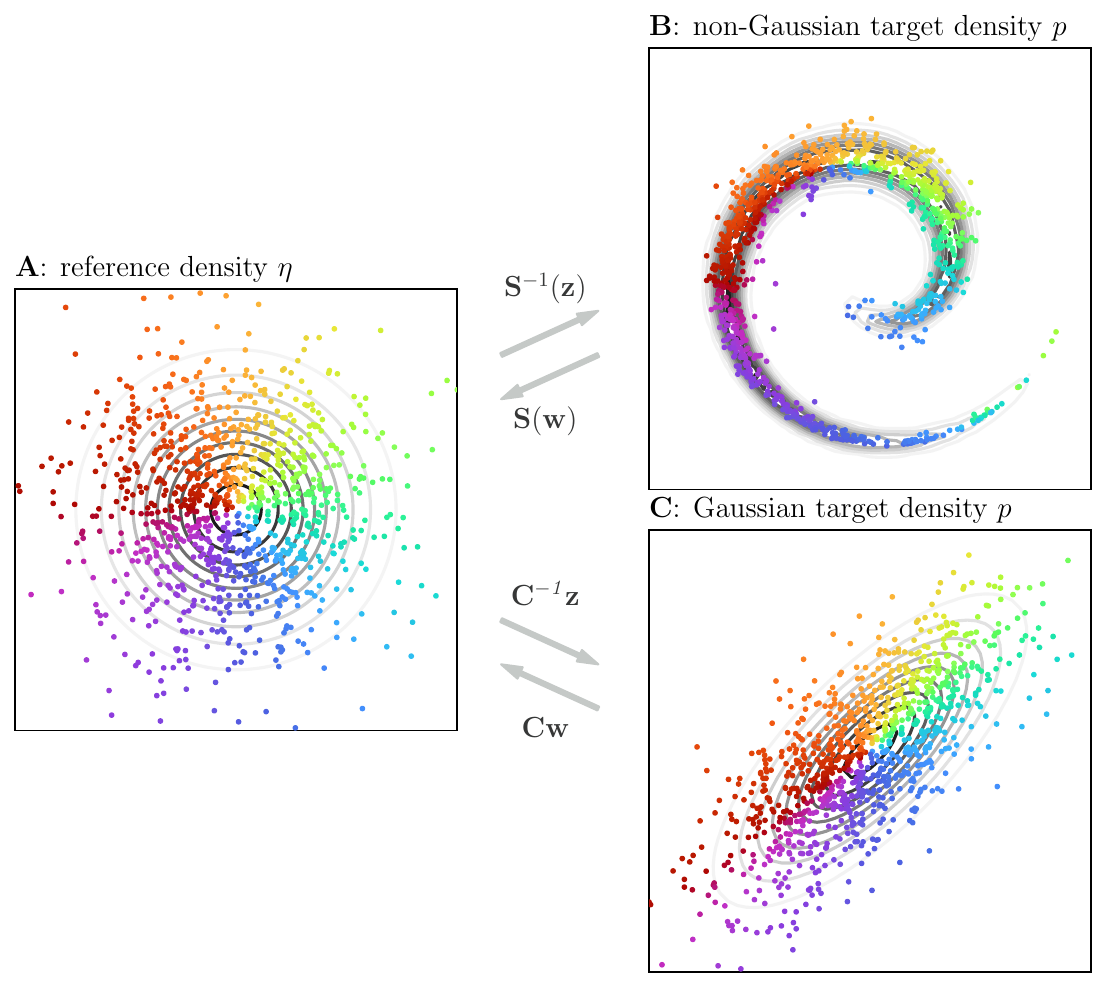}
  \caption{With the correct triangular transport map $\SKR$, we can convert samples $\w \in \mathbb{R}^{K}$ from  an arbitrary target distribution $p$ (panels B and C) into samples $\z \in \mathbb{R}^{K}$ from a multivariate standard Gaussian reference distribution $\eta$ (panel A). If $p$ is non-Gaussian (panel B), we need nonlinear maps. If $p$ is Gaussian (panel C), a linear lower-triangular map $\SKR(\w) = \mathbf{C}\w$, for $\mathbf{C} \in\mathbb{R}^{K \times K}$, suffices. Grey lines denote density contours, and colored scatter points indicate where reference samples are mapped to and from.}
  \label{fig:transport_map}
\end{figure}

Transport methods seek transformations between random variables and hence probability distributions, and can be used to sample from the conditional distributions that arise in Bayesian inference problems. In general, two probability density functions (PDFs) of interest can be related by the \textit{change-of-variables} formula. Here, this formula is used to approximate a target pdf $p(\w)$ which characterizes a random variable of interest $\w$. This target pdf is related via an invertible transformation $\SKR$ to another random variable $\z$ with a known ``reference'' pdf $\eta$ \citep[e.g.,][]{Villani2007OptimalNew, Santambrogio2015OptimalMathematicians} (Figure~\ref{fig:transport_map}):
\begin{equation}
p(\w) \approx \SKR ^{\sharp}\eta(\w) \coloneqq \eta(\SKR (\w))\det \mathbf{\nabla} \SKR (\w).
\label{eq:change_of_variables}
\end{equation}
The \textit{target} PDF $p(\w)$ is approximated by a {\it pullback pdf} $\SKR ^{\sharp}\eta$ that depends on the \textit{reference} PDF $\eta(\z)$, evaluated at $\z=\SKR (\w)$. Here we assume that both $p$ and $\eta$ are pdfs on $\mathbb{R}^K$. The transport map is exact if and only if Equation~\ref{eq:change_of_variables} is a strict equality; otherwise the pullback pdf is an approximation to $p$. The term $\det {\nabla} \SKR (\w)$ compensates for the differential change in volume between the $\w$ and $\z$ coordinate systems, induced by the map $\SKR $. Since the transformation is invertible, an analogous change of variables formula approximates the reference pdf $\eta(\z)$ via a \textit{pushforward pdf} $\SKR _{\sharp}p$ that depends on $p(\w)$, evaluated at $\w=\SKR^{-1}(\z)$.

The invertible \textit{transport map} $\SKR$ for a particular inference problem is selected to give a pullback $\SKR^{\sharp}\eta$ that adequately approximates the desired target distribution $p$. An efficient way to construct such a transport map in $K>1$ dimensions is via the \textit{Knothe--Rosenblatt} (KR) \textit{rearrangement} \citep{Villani2007OptimalNew,Rosenblatt1952RemarksTransformation}:
\begin{equation}
\SKR (\w)=\begin{bmatrix*}[l]
    S_{1}(w_{1}) \\
    S_{2}(w_{1},w_{2}) \\
    ~~ \vdots \\
    S_{K}(w_{1},\dots,w_{K})
\end{bmatrix*}=\begin{bmatrix*}[c]
    z_{1} \\
    z_{2} \\
    \vdots \\
    z_{K}
\end{bmatrix*} = \z
\label{eq:decomposed_S}.
\end{equation}
The KR map $\SKR$ is lower triangular, in that it comprises $K$ component functions $S_{k}\colon \mathbb{R}^k \to \mathbb{R}$, each of which depends at most on the first $k$ coordinates of $\w$ and is monotone in the $k$-th coordinate ($\partial_{w_{k}}S_{k}>0$)~\citep{Santambrogio2015OptimalMathematicians}. This structure has a number of useful properties. First, a triangular KR map between any absolutely continuous $p$ and $\eta$ exists and is unique (up to ordering)~\citep{Villani2007OptimalNew}. The monotonicity property just described guarantees invertibility, and the determinant of map's Jacobian in Equation~\ref{eq:change_of_variables} can be evaluated efficiently as the product of its diagonal entries (i.e., $\det\nabla \SKR =\prod_{k=1}^{K}\partial_{w_{k}}S_{k}$).

While the components $S_{k}$ of the forward map $\SKR$ can be evaluated independently and in parallel at any input $\w$, evaluating the map inverse requires solving a sequence of one-dimensional root finding problems starting from the top of the map ($S_{1}$) and moving downwards. 
Let $S_{k}^{-1}(w_1,\dots,w_{k-1};  \cdot \, )$ denote the inverse of the function $x \mapsto S_{k}(w_{1},\dots,w_{k-1},x)$. Then, the inverse of $\SKR$ in Equation~\ref{eq:decomposed_S} is given by 
\begin{equation}
\SKR^{-1}(\z)=\begin{bmatrix*}[l]
    S_{1}^{-1}(z_{1}) \\
    S_{2}^{-1}(w_{1}; z_{2}) \\
    ~ ~ \vdots \\
    S_{K}^{-1}(w_{1},\dots,w_{K-1}; z_{K})\end{bmatrix*}=\begin{bmatrix*}[c]
    w_{1} \\
    w_{2} \\
    \vdots \\
    w_{K}
\end{bmatrix*}  = \w . 
\label{eq:inverse_decomposed_S}
\end{equation}
In each row of $\SKR^{-1}$, the previous inversions' outputs $w_{1},\dots ,w_{k-1}$ serve as input for the next inversion $S_{k}^{-1}$. 

An important and extremely useful consequence of this dependence on the output of previous entries is that the map component inverses $S_{k}^{-1}$ sample certain \textit{marginal conditionals} of the target pdf $p$, according to a factorization that reflects the chosen variable ordering. Recall that any target pdf $p(\w)$ can be factorized in the following telescoping way:
\begin{equation}
p(\w ) = p(w_{1})p(w_{2}|w_{1})p(w_{3}|w_{1},w_{2})\cdots p(w_{K}|w_{1},\dots,w_{K-1}).
\label{eq:marginal_conditional_telescopic}
\end{equation}
This factorization expresses $p(\w)$ as a product of its marginal conditional densities. Evaluating $\SKR^{-1}$ on a sample $\z^i$ drawn from the reference distribution $\eta$ produces a sample $\w^i$ from the target pdf $p$ by sequentially sampling each of these marginal conditional densities. Specifically, each row of Equation~\ref{eq:inverse_decomposed_S} samples the $k$th component of $\w^i$, $w_k^i$, conditioned on the realization of the previous $k-1$ components $\w^i_{1:k-1}$:
\begin{equation}
w_{k}^{i} = S_{k}^{-1}(w_{1}^{i},\dots,w_{k-1}^{i}; z_{k}^{i}) \sim S_{k}^{-1}( w_{1}^{i},\dots,w_{k-1}^{i}; \cdot \, )_\sharp \eta_{k}  =  p (w_{k} \vert w^i_{1},\dots,w^i_{k-1} ),
\label{eq:marginal_conditional}
\end{equation}
where $\eta_k$ is the $k$th marginal of the reference pdf. Note that Equation~\ref{eq:marginal_conditional} only applies if the reference random variables are mutually independent or, equivalently, if the reference pdf $\eta(\z)$ can be written as the product of its scalar marginals. A standard Gaussian reference fulfills this property. The construction in Equation~\ref{eq:marginal_conditional} shows the direct link between triangular maps and conditioning, where each map component function $S_k$ corresponds to a specific marginal conditional factor of $p$. 

\begin{figure}
  \centering
  \includegraphics[width=\textwidth]{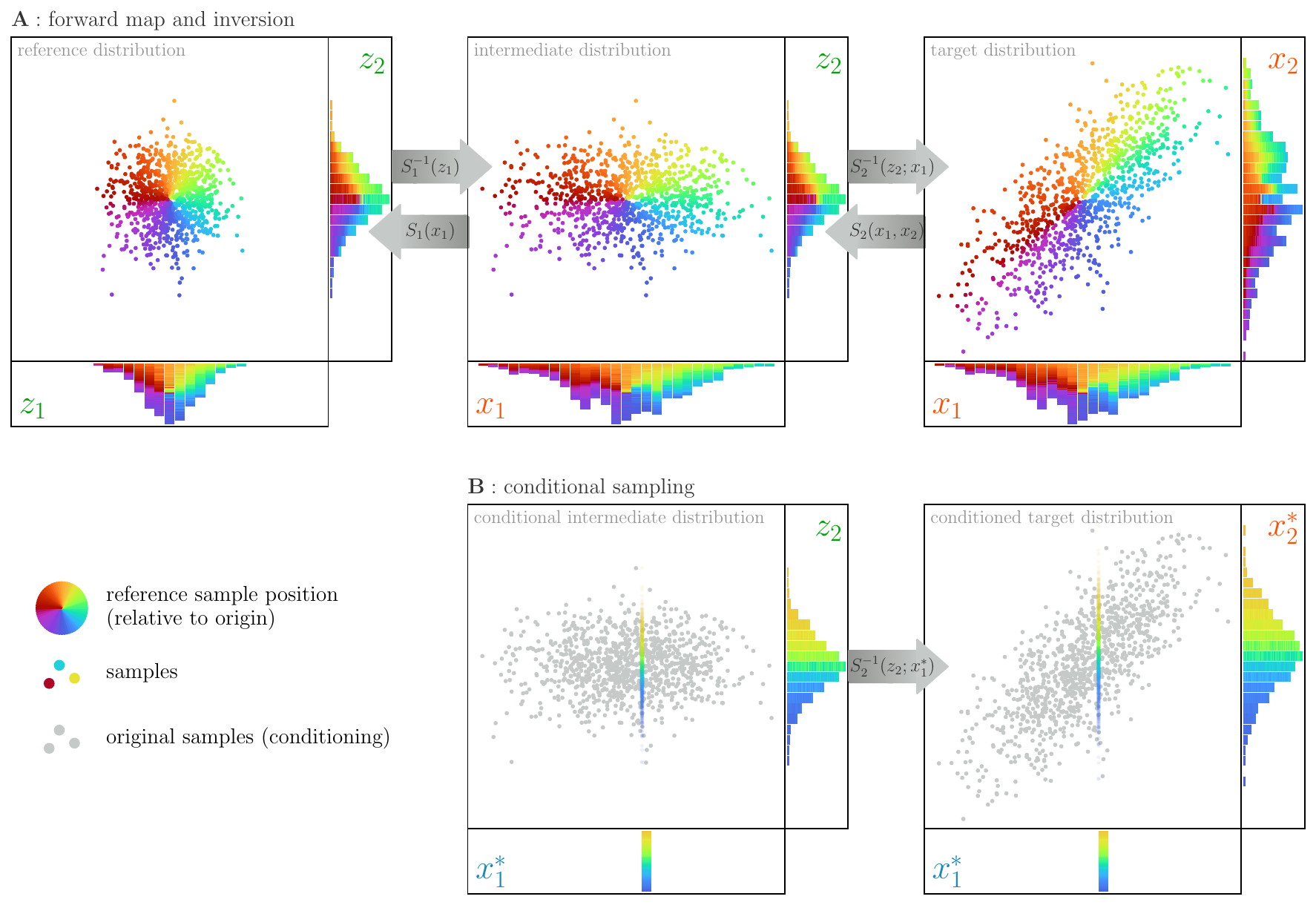}
  \caption{(A) The forward map (top row, from right) and its inverse (top row, from left) operate via an implicit intermediate distribution (center). Each map component only transforms one of the coordinates. (B) Supplying the inverse map with a modified input distribution (bottom row)---specifically, fixing $x_1$ to  $x_1^*$---yields samples from a specific conditional of the target distribution.}
  \label{fig:conditioning}
\end{figure}

Overall, this discussion underscores three important properties of triangular transport maps \citep{Spantini2018InferenceCouplings}: 
\begin{enumerate}
    \item \textbf{Importance of ordering}: The telescoping decomposition of $p(\w)$ (see Equation~\ref{eq:marginal_conditional_telescopic}) can be taken in an arbitrary order, i.e., there is a different telescoping decomposition for any permutation of the entries of $\w$ in $\SKR^{-1}$ (see Equation~\ref{eq:inverse_decomposed_S} and Equation~\ref{eq:marginal_conditional}). As we shall see, this choice has important and useful implications for inference problems and, in particular, smoothing.
    \item \textbf{Realizing conditional independence}: In many systems, we can \textit{sparsify} the factorization in Equation~\ref{eq:marginal_conditional_telescopic} by exploiting conditional independence. Triangular transport methods can realize this effect by removing the corresponding variable dependencies from the transport map components: For instance, if $w_3$ is conditionally independent of $w_2$ given $w_1$, i.e. if $p(w_3|w_1,w_2) = p(w_3|w_1)$, then $S_3(w_1,w_2, w_3) = S_3(w_1, w_3)$; i.e., the third component function of the KR map $\SKR$ does not depend on $w_2$. (As above, this requires that the reference $\eta$ be a product distribution; see \citet[Theorem 3]{Spantini2018InferenceCouplings} for a complete elucidation of this relationship.)  As a result, $S_{3}^{-1}(w_1; z_3)$ also samples exactly from the marginal conditional.
    \item \textbf{Bayesian inference}: Triangular transport methods can not only sample the target $p$, but also its conditionals. This can be achieved by skipping the upper map component inversions $S_{j}^{-1}$ for $1 \leq j < k$, and replacing their corresponding outputs with the desired values $w_{j}^{*}$ for the conditioning variables. Inverting the remaining map components then yields samples from the conditional $p(\w_{k:K}|\w_{1:k-1}^{*})$. This process is illustrated in Figure~\ref{fig:conditioning}.
\end{enumerate}

\subsection{Learning and parameterizing maps}
\label{sec:mapparam}
So far, we have kept $p$ and $\eta$ general and described properties of the exact KR map $\SKR$, without delving into implementation-related details. In practice, we are interested in approximating the target distribution $p$, which in general may be accessible only via (i) unnormalized density evaluations, or (ii) a set of samples. In the ensemble filtering/smoothing setting, the latter case is of interest. Given a set of samples from the target, we wish to identify a map $\SKR$ that pushes forward this target to a well-defined reference distribution, usually a multivariate standard normal distribution $\eta=\mathcal{N}(\mathbf{0},\mathbf{I})$. Once we have identified this map, we can freely transform samples from the reference to the target, and back; see Figure~\ref{fig:transport_map}. This map allows us to sample from the target distribution by sampling the reference $\z^i \sim \eta$, and transforming each reference sample $\z^i$ into a target sample $\w^i = \SKR^{-1}(\z^i)$ through the inverse map. 

To find this map, we follow \citet{Marzouk2017SamplingIntroduction} and minimize the Kullback--Leibler divergence from $\SKR^{\sharp}\eta$ to $p$ over some \textit{parameterized family} of monotone lower triangular functions $\mathcal{S}$. We note that it is possible to base this optimization objective on other divergences. However, the forward Kullback--Leibler divergence is natural because it corresponds to maximum likelihood estimation, once the expectations are discretized with a finite set of samples. Functions in the chosen family can be highly nonlinear---although we will focus on linear functions in this paper, leaving more general nonlinear maps to the companion paper~\citep{Ramgraber2022underUpdatesb}. The exposition below remains general, however, so that it encompasses any choice of map parameterization.

Using Equations~\ref{eq:change_of_variables} and \ref{eq:decomposed_S}, as well as a Monte Carlo approximation of $p$, we can derive the objective functions
\begin{equation}
\widehat{\mathcal{J}}_{k}(S_{k}) = \sum_{i=1}^{N}\left(\frac{1}{2}S_{k}(\w^{i})^{2} - \log\frac{\partial S_{k}(\w^{i})}{\partial w_{k}} \right), \ k=1, \dots, K
\label{eq:objective_function}
\end{equation}
where the superscript $i=1,\dots ,N$ denotes sample indices from an ensemble $\{\w^{i}\}_{i=1}^{N}\sim p$. Minimizing each objective $\widehat{\mathcal{J}}_k$ over some family of functions from $\mathbb{R}^k \to \mathbb{R}$ that are monotone increasing in $x_k$, which we denote by $\mathcal{S}_k$, yields an estimate $\widehat{S}_k$ for each map component $S_k$. The derivation of Equation~\ref{eq:objective_function} is provided in Appendix~\ref{sec:AppendixA}. This objective function has a convenient intuitive interpretation. Minimization of $\widehat{\mathcal{J}}_{k}$ with respect to the map component $S_k$ attempts to drive the first term within the summation to zero by mapping the samples $\w^{i}\sim p$ to values as close to zero (the mode of $\eta$) as possible. By contrast, it attempts to drive the term after the minus sign to infinity by maximizing the map's derivative values, which spreads out the samples as much as possible. The optimal compromise between these two antagonistic objectives depends on the sample values and on constraints imposed by the parameterization adopted for the map. 

The most effective way to minimize each objective in Equation~\ref{eq:objective_function} over a set of candidate maps $S_k \in \mathcal{S}_k$ depends very much on the map parameterization, i.e., the choice of spaces $(\mathcal{S}_k)_{k=1}^K$. In general, the objective $\widehat{\mathcal{J}}_k$ is a convex function of the map $S_k$. Nonlinear parameterizations of $S_k$, introduced to ensure monotonicity in general settings, typically break this convexity, but certain attractive properties of the optimization problem might remain; for details, see \citet{Baptista2020OnMaps}. More restrictive/specific parameterizations of nonlinear functions $S_k$, including linear parameterizations of such functions, can enable efficient and often closed-form optimization solutions; see \citet[Appendix A]{Spantini2022CouplingFiltering}. We will return to these issues in the companion paper \citet{Ramgraber2022underUpdatesb}, where we employ nonlinear transport maps $S_k$. In the numerical experiments of the present paper, we restrict our attention to affine transport maps; that is, we choose the spaces $\mathcal{S}_k$ to consist only of affine functions on $\mathbb{R}^k$. In this setting, a minimizer of each objective in Equation~\ref{eq:objective_function} can be written in closed form, as we shall describe in Section~\ref{subsec:linear_maps}.

Sparsity of the KR map also has implications on the solution of the optimization problem above. In particular, if the conditional independence structure of the target distribution $p$ guarantees that certain component functions $S_k$ of the map do not depend on certain input variables (e.g., some subset of $\w_{<k}$), this sparsity can be imposed on the optimization problem by simplifying the chosen approximation space $\mathcal{S}_k$. In other words, we need not search over functions that have superfluous variable dependence. Doing so can mitigate the curse of dimensionality and vastly reduce the variance of the estimated maps $\widehat{S}_k$; see \citet{Morrison2017BeyondSetting}. Note that the complexity of representing or storing a dense linear map with $K$ components scales as $\mathcal{O}(K^2)$, while for a separable parameterization of a nonlinear map (see Section 3.3.2 of \citet{Ramgraber2022underUpdatesb}) this becomes $\mathcal{O}(MK^2)$, where $M$ represents the maximum polynomial degree or a similar notion of basis size. On the other hand, a suitably sparse map may scale linearly with the dimension $K$, i.e., as $\mathcal{O}(MKD)$, if each map component $S_{k}$ depends at most on $D$ neighbors of $x_k$.
Imposing such sparsity is also an important way of regularizing the map learning problem and achieving localization \citep{Spantini2022CouplingFiltering}; we will revisit this topic below and in \citet{Ramgraber2022underUpdatesb}. 

\subsection{Block conditional sampling}
\label{sec:condsamp}
We note that it is also possible, and often useful, to arrange Equations~\ref{eq:decomposed_S} and~\ref{eq:inverse_decomposed_S} in block form. For example, if we have a input vector of the form $\w=[\y,\x]^\top =[y_1,\dots,y_M,x_1,\dots,x_D]^\top $, we can divide the corresponding transport map into two blocks as follows:
\begin{equation}
\SKR(\y,\x)=\left[\begin{array}{lr}
    S_{1}(y_{1}) \\
    \vdots \\
    S_{M}(y_{1},\dots,y_{M}) \\[3pt]
    \hline \vspace{3pt}
    S_{M+1}(y_{1},\dots,y_{M},x_{1}) \\
    \vdots \\
    S_{M+D}(y_{1},\dots,y_{M},x_{1},\dots,x_{D})
\end{array}\right]=\left[\begin{array}{lr}
    \SKR_{\y}(\y) \\[2pt]
    \SKR_{\x}(\y,\x)
\end{array}\right] ,
\label{eq:decomposed_S_block_form}
\end{equation}
where we have placed the elements of the observation vector $\y$ in the upper block of the map, so that the corresponding factorization of the target distribution (Equation~\ref{eq:marginal_conditional_telescopic}) is $p(\y,\x) = p(\y)p(\x|\y)$. The horizontal line in Equation~\ref{eq:decomposed_S_block_form} is a visual aid to distinguish the two map blocks. This block arrangement enables us to characterize posterior distributions for $\x$ conditioned on specific observations $\y^{*}$, such as $p(\x|\y^{*})$, by setting $\y=\y^{*}$. The lower block of the transport map, $\SKR_{\x}$, can then be used to draw samples from $p(\x|\y^{*})$ by inverting the function $\x \mapsto \SKR_{\x}(\y^{*}, \x)$ (recall the discussion around Equation~\ref{eq:marginal_conditional}). 

In fact, one can consider \textit{two} possible methods for conditional sampling. If we wish to sample the posterior $p(\x|\y^{*})$, the approach described above involves drawing $D$-dimensional standard Gaussian samples $\Z \sim \eta_D$ from the reference marginal, and then evaluating them at the inverse of $\x \mapsto \SKR_{\x}(\y^{*}, \x)$, which we denote by $\SKR_{\x}^{-1}(\y^{*}; \Z)$. In practice, however, if there is any error in the estimated map, it is better to use (approximate) reference samples obtained from an application of the forward map $\SKR_{\x}(\Y,\X)$. This equation can be composed with $\SKR_{\x}^{-1}(\y^{*}; \cdot \, )$ to yield a \textit{composite map} $\T _{\y^{*}}$ which conditions samples from the joint distribution on a specified value $\y^{*}$ of $\y$:
\begin{equation}
\x^{*} = 
\T_{\y^{*}}(\y,\x) \coloneqq 
\SKR_{\x}^{-1}(\y^{*}; \cdot)\circ\SKR_{\x}(\y,\x).
\label{eq:composite_map}
\end{equation}
This composite map first applies the lower block of the forward map $\SKR_{\x}(\Y,\X)$ to generate pushforward samples $\widetilde{\Z} \sim \left (\SKR_{\x} \right )_{\sharp} p$, the map's approximation to the reference marginal $\eta_D$. These samples then serve as input for the map's inversion to obtain $\X^{*} = \SKR_{\x}^{-1}(\y^{*}; \widetilde{\Z})$. As described in \citet{Spantini2022CouplingFiltering}, composite maps often allow us to use simpler maps than otherwise necessary. Approximation errors are cancelled between an imperfect map $\SKR$ and its (partial) inverse in Equation \ref{eq:composite_map}. By using samples $(\Y, \X) \sim p$ as inputs, rather than standard Gaussian samples $\Z \sim \eta_D$, this procedure retains features of the target distribution that the map failed to capture. If, on the other hand, we were to generate target samples from an imperfect map via true Gaussian reference samples, we would only sample conditionals of the (imperfect) pullback distribution $\SKR^{\sharp}\eta$, i.e., the map's approximation to the target $p$. This same preservation of uncaptured features also underlies the EnKF's surprising efficiency in non-Gaussian settings, as we shall see in the next section.

\subsection{Linear transport maps}\label{subsec:linear_maps}

Let us now consider an illustrative example in which we have an arbitrary $K$-dimensional multivariate Gaussian target distribution $p=\mathcal{N}(\boldsymbol{\mu},\boldsymbol{\Sigma})$ with mean vector $\boldsymbol{\mu}$ and covariance matrix $\boldsymbol{\Sigma}$. Without loss of generality, we will assume $\boldsymbol{\mu}=\mathbf{0}$ in the following, as we can standardize the target ensemble before applying the transport map, as long as we undo this shift afterwards. We wish to find a lower-triangular transport map $\SKR$ which transforms samples $\w \sim p$ into samples $\z \sim \eta$ from the standard  Gaussian reference distribution $\eta=\mathcal{N}(\mathbf{0},\mathbf{I})$. In this Gaussian setting, the transport map is necessarily linear, and hence our lower-triangular $\SKR\left(\w\right)$ can be expressed as a product of a lower-triangular coefficient matrix $\mathbf{C} \in \mathbb{R}^{K \times K}$ with the vector $\w$:
\begin{equation}
    \SKR(\w) = \mathbf{C}\w =
    \begin{bmatrix*}[c]
        c_{1,1}     \\
        c_{2,1}     && c_{2,2}      \\
        c_{3,1}     && c_{3,2}      && c_{3,3}      \\
        \vdots      && \vdots       && \vdots       && \ddots   \\
        c_{K,1}     && c_{K,2}      && c_{K,3}      && \hdots   && c_{K,K}
    \end{bmatrix*} \begin{bmatrix*}[c]
        w_{1}      \\
        w_{2}      \\
        w_{3}      \\
        \vdots      \\
        w_{K}
    \end{bmatrix*}=\z.
    \label{eq:linear_map}
\end{equation}

With access to the covariance matrix $\boldsymbol{\Sigma}$, the coefficient matrix $\mathbf{C}$ that minimizes the KL divergence from $\SKR^\sharp\eta$ to $p$ can be derived in closed form. It suffices to recognize that the covariance of the Gaussian pushforward pdf $\SKR_{\sharp} p$ should match the covariance of the reference pdf under this choice. That is,
\begin{equation}
    \mathbf{I} = \mathbb{E}[\z\z^\top ] = \mathbb{E} [\left(\mathbf{C}\w\right)\left(\mathbf{C}\w\right)^\top ] = \mathbf{C} \, \mathbb{E}[\w\w^\top]\mathbf{C}^\top  = \mathbf{C}\boldsymbol{\Sigma}\mathbf{C}^\top ,
    \label{eq:pre_cholesky}
\end{equation}
where $\mathbb{E}[\z]=\mathbf{0}$ and $\mathbb{E}[\w]=\mathbf{0}$ as assumed above. Writing the Cholesky factorization of the covariance matrix $\boldsymbol{\Sigma} = \mathbf{L}\mathbf{L}^{\top }$ shows that the lower triangular coefficient matrix $\mathbf{C}$ is simply the inverse of the Cholesky factor $\mathbf{L}$, i.e., $\mathbf{C}=\mathbf{L}^{-1}$. It is also important to note that the precision matrix $\boldsymbol{\Sigma}^{-1}$ of a multivariate Gaussian encodes all pairwise conditional independence properties of the components of $\w$. Manipulating Equation~\ref{eq:pre_cholesky} shows that the precision matrix is connected to the triangular map coefficient matrix $\mathbf{C}$ via $\boldsymbol{\Sigma}^{-1} = \mathbf{C}^{\top }\mathbf{C}$. In other words, $\mathbf{C}$ can be understood as a Cholesky factor of the precision matrix (in the Gaussian case), and thus it inherits sparsity when the precision matrix is sparse---i.e., when the target distribution has conditional independence. For more on this topic, see \cite{Baptista2021LearningTransport,Morrison2017BeyondSetting,Spantini2018InferenceCouplings}. 

An example of the resulting linear transformation is illustrated in Figure~\ref{fig:transport_map}. Adopting the decomposition $\w=[\y,\x]^\top $ once more, it can be demonstrated that the composite transport map $\T $ in Equation~\ref{eq:composite_map}, when setting $\SKR = \mathbf{C}\w$, retrieves the equation for conditioning a sample drawn from a multivariate Gaussian distribution (see Appendix~\ref{sec:AppendixB} for a derivation):
\begin{equation}
    \x^{*} = \T _{\y^{*}}(\y,\x) = \x - \boldsymbol{\Sigma}_{\x,\y}\boldsymbol{\Sigma}_{\y,\y}^{-1}\left(\y - \y^{*}\right).
    \label{eq:conditioning_Gaussian_samples}
\end{equation}
In other words, this linear map $\T _{\y^{*}}$ transforms jointly Gaussian samples $[\y, \x]$ to conditional Gaussian samples $\x^{*} \sim p(\x \vert \y^{*}).$ See also \citet{lehtinen1989linear} for a derivation of this transformation in a generalized functional setting.

While we derived the maps $\SKR$ and $\T _{\y^{*}}$ above under the assumption that the target distribution $p$ was multivariate Gaussian, it is important to note that these formulas are in fact {much more general}. If the transport map approximation spaces $(\mathcal{S}_k)_{k=1}^K$ are restricted to consist only of affine functions, then these formulas hold for \textit{any} generic {non-Gaussian} target distribution $p(\w)$ (so long as it has finite mean and variance) with standard Gaussian reference $\eta$. More precisely, minimizers of the $K$ objective functions $\widehat{\mathcal{J}}_k$ in Equation~\ref{eq:objective_function}, over affine functions, can be written in closed form as follows: given an ensemble $\{\w^{i}\}_{i=1}^{N}\sim p$, the optimal lower-triangular affine map is $\widehat{\SKR }(\w) = (\widehat{S}_1(w_1), \widehat{S}_2(w_{1:2}), \ldots, \widehat{S}_K(w_{1:K}))^\top  = \widehat{\mathbf{C}} (\w - \widehat{\boldsymbol{\mu}} )$, where $\widehat{\boldsymbol{\mu}}$ is the ensemble mean and $\widehat{\mathbf{C}}$ is the inverse of the Cholesky factor of the ensemble covariance matrix, i.e., $\widehat{\mathbf{C}} = \widehat{\mathbf{L}}^{-1}$ with $\widehat{\boldsymbol{\Sigma}}_{\w,\w} = \widehat{\mathbf{L}} \, \widehat{\mathbf{L}}^{\top}$. A derivation can be found in the proof of Proposition 8 in~\citet{Baptista2021LearningTransport}. 
Some intuition for this result comes from recalling that minimization of the forward Kullback--Leibler divergence $\KLDiv(p||\SKR ^\sharp \eta)$ over its second argument (which led to the objective in Equation~\ref{eq:objective_function}) seeks \textit{moment matching}, and that the restriction to affine maps allows $\SKR ^\sharp \eta$ to be an arbitrary multivariate Gaussian; hence the best such choice matches the first and second (empirical) moments from $p$. 

Introducing again the decomposition $\w=[\y,\x]^\top $, the affine lower-triangular map $\widehat{\SKR }$ described in the paragraph above leads to precisely the same \textit{composite map} as in Equation~\ref{eq:conditioning_Gaussian_samples}, but with the two covariance matrices replaced by their ensemble estimates $\widehat{\boldsymbol{\Sigma}}_{\x,\y}$ and $\widehat{\boldsymbol{\Sigma}}_{\y,\y}$. This transformation performs consistent conditioning on $\y^{*}$ when $\x$ and $\y$ are jointly Gaussian (as $N \rightarrow \infty$), and is a linear approximation to the conditioning operation otherwise. Equation~\ref{eq:conditioning_Gaussian_samples}, or its ensemble approximation, lies at the heart of all Kalman-type filters and smoothers. We note that the original paper of \citet{kalman1960kalmanfilter} arrived at this update by seeking the minimum mean-square error estimator of the state $\x$ among all affine functions of the observations $\y$. Kalman's estimator also uses second moments of $\x$ and $\y$ and holds without any assumptions of Gaussianity. One subtle distinction is that Equation~\ref{eq:conditioning_Gaussian_samples} is not cast as a point estimator of the state, but rather as an update of the prior random variable $\x$ itself; yet it yields the same estimate as Kalman via its mean.
In this study, we will keep Equation~\ref{eq:conditioning_Gaussian_samples} as general as possible, positing no further assumptions on the nature of the observation and forecast errors. 

\section{Ensemble transport smoothers (EnTS)} \label{sec:ents_all}
Having established the use of transport maps for Bayesian inference in Section~\ref{sec:transport_maps}, we proceed to discuss their application to smoothing. Recursive smoothing algorithms exploit the Markov structure described in Figure~\ref{fig:Markovian_graph} by processing measurements sequentially, at each time step in the smoothing interval, rather than by processing the entire set of measurements simultaneously~\citep{Evensen2000AnDynamics}. This enables us to derive and apply a separate transport map at each update/analysis step in the recursive smoothing algorithm. Since each update step works only with the states and measurements associated with a particular update time, the dimensionality of the considered map is much smaller than it would be if all measurements were processed at once. 

To examine this further, assume we have samples $(\Y_{t},\X_{1:t})$ from the distribution $p(\y_{t},\x_{1:t}|\y_{1:t-1}^{*})$ at some generic time $t$. This distribution encapsulates prior information about all states through time $t$ and about the measurement at time $t$, given measurements obtained through time $t-1$. 
If we have samples from the previous smoothing distribution $p(\x_{1:t-1}|\y_{1:t-1}^{*})$, we can generate samples from this ``extended'' forecast $p(\y_{t},\x_{1:t}|\y_{1:t-1}^{*})$ by simulating the dynamics $p(\x_t \vert \x^i_{t-1})$ for each ensemble member $\x_{t-1}^i$, then sampling from the observation model $p(\y_t \vert \x_t^i)$ given the corresponding $\x_t^i$, and finally appending these realizations to each $\x_{1:t-1}^i$.

The update operation at time $t$ seeks to transform this forecast distribution to reflect conditioning on a new observation value $\y_{t}^{*}$, thus producing samples $\X_{1:t}^{*}$ from the updated smoothing distribution $p(\x_{1:t}|\y_{1:t}^{*})$. There are a number of ways to use transport maps to carry out this smoothing update, depending on how we order the variables within the generic triangular transport map $\SKR(\w)$ of Equation~\ref{eq:decomposed_S}. Following the approach taken in Equation~\ref{eq:decomposed_S_block_form}, we partition the inputs of the map into a block of observations $\y_t$ above and a block of states $\x_{1:t}$ below. Section~\ref{sec:any_ordering} presents the resulting update in the most generic case---with an arbitrary ordering of the individual states within the (lower) state block and hence no exploitation of the conditional independence relationships among the states. This update yields the ``dense'' nonlinear ensemble transport smoother (EnTS).

\begin{figure}[!ht]
  \centering
  \includegraphics[width=\textwidth]{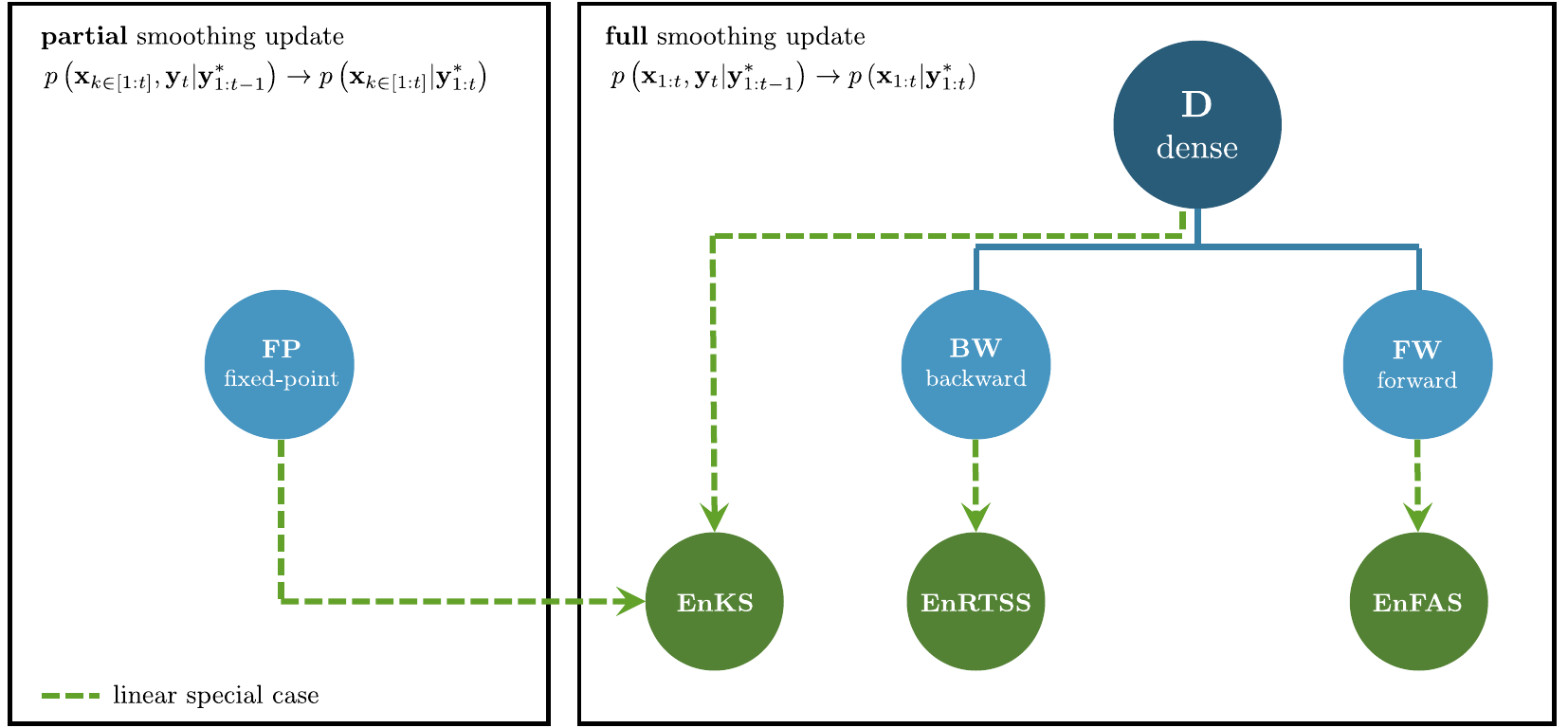}
  \caption{A taxonomy of the smoothers considered in this paper. Backward and forward smoothers arise as sparse variants of a generic, dense EnTS. If the maps are constrained to be affine functions, these smoothers yield the EnRTSS or EnFAS as special cases. The EnKS emerges as a linear special case of either a dense smoother or a sparse fixed-point smoother.}
  \label{fig:smoother_taxonomy}
\end{figure}

Next, we show how specific orderings of the state variables can lead to more effective algorithms that exploit \textit{sparse variable dependence} in the triangular transport map (see Figure~\ref{fig:sparsity_patterns}). Sections~\ref{sec:ordering_A} and~\ref{sec:ordering_B} present smoothing algorithms based on two different orderings for the state variables. Section~\ref{subsec:fixedpoint} then presents a specialized algorithm for fixed-point smoothing. In Sections~\ref{sec:any_ordering}, \ref{sec:ordering_A}, and \ref{subsec:fixedpoint} we also elucidate relationships between these transport map smoothers and well-known Kalman-type smoothing recursions. Connections among all these smoothing algorithms are summarized in Figure~\ref{fig:smoother_taxonomy}. 

Section~\ref{subsec:subsequent_iterations} discusses how to carry out subsequent iterations of these recursive smoothers in an online setting (i.e., to process the next observation $\y^{*}_{t+1}$, and so on). Pseudo-code for the proposed ensemble transport smoothing algorithms is provided in Appendix~\ref{sec:AppendixC}.

\subsection{Joint update with a dense map}\label{sec:any_ordering}

First we consider the conceptually simplest approach to a smoothing update. Recalling the construction in Section~\ref{sec:condsamp}, we set $\y = \y_t$ and $\x = \x_{1:t}$ in Equation~\ref{eq:decomposed_S_block_form}. For each value of $\y_t$, the lower block of the map is a triangular function, $\x_{1:t} \mapsto \SKR_{\x_{1:t}}(\y_t,\x_{1:t})$. For a generic permutation of $\{1,\dots,t\}$, the triangular map will not reflect conditional independence among the states in general,
and thus each component function of this map will depend on all state variables preceding it in the chosen ordering, as illustrated in Figure~\ref{fig:sparsity_patterns}A. (We allow the ordering of states to be arbitrary here; the next sections will consider specific orderings that exploit the Markovian structure of the state sequence and yield sparse variable dependence in the map.) The resulting composite map that updates the states is precisely as in Equation~\ref{eq:composite_map}:
\begin{equation} \label{eq:composite_map_dense_update}
    \x_{1:t}^* = \T_{\y_t^*}(\y_t, \x_{1:t}) = \SKR_{\x_{1:t}}^{-1}(\y_t^*,\cdot) \circ \SKR_{\x_{1:t}}(\y,\x_{1:t}).
\end{equation}
This composite map updates the entire collection of states $\x_{1:t}$ all-at-once. We note that the map $\SKR_{\x_{1:t}}$ is estimated from the joint ensemble $(\Y_t, \X_{1:t})$. 

\subsubsection{Relation to the ensemble Kalman smoother}
\label{sec:densetoEnKS}

As might be anticipated following the discussion in Section~\ref{subsec:linear_maps}, restricting a transport smoother to employ only affine maps reproduces classical Kalman-type smoothing algorithms. \citet{Spantini2022CouplingFiltering} established this connection to the EnKF, for filtering. We will see a similar correspondence here, and for many of the smoothing constructions described in subsequent sections.

Focusing on the dense ensemble transport smoother (EnTS) introduced above, suppose we restrict the map $\SKR_{\x_{1:t}}(\y_t, \x_{1:t})$ above to be an affine function, i.e., a function of the form
$$
\SKR_{\x_{1:t}}(\y_t, \x_{1:t}) = \mathbf{C} \begin{bmatrix} \y_{t} \\ \x_{1:t} \end{bmatrix} + \mathbf{b},
$$
where $\mathbf{C}$ is a rectangular coefficient matrix whose row dimension equals $t$ times the dimension of the state vector. Minimizing the objective in Equation~\ref{eq:objective_function} over maps of this form yields a $\mathbf{C}$ that is the lower block of the inverse Cholesky factor of the \textit{joint covariance matrix} of $(\y_t,\x_{1:t})$, as described in Section~\ref{subsec:linear_maps}. The constant $\mathbf{b} \in \R^{d}$ captures the ensemble means. Hence, elements of $\mathbf{C}$ are based on  cross-covariances between pairs of states and between the states and the observation; see Appendix~\ref{sec:AppendixB}. Using this $\SKR_{\x_{1:t}}$, we can then construct a composite map. Exactly as in Equation~\ref{eq:conditioning_Gaussian_samples}, this composite map reduces to
\begin{equation}
\x_{1:t}^* = \x_{1:t} - \bfSigma_{\x_{1:t},\y_t} \bfSigma_{\y_t}^{-1} (\y_t - \y_t^{\ast}),
\label{eq:firstEnKS}
\end{equation}
which is indeed the well-known ensemble Kalman smoother (EnKS).

As compared to the full map $\SKR$, the EnKS update above only involves a subset of the elements of the coefficient matrix. In particular, it only requires knowing the cross-covariance between each state $\x_s$, $s=1,\dots, t$, and the observation $\y_t$. Furthermore, the joint update in Equation~\ref{eq:firstEnKS} trivially decomposes into an independent update for each state $\x_s$; each update is captured by a separate row-block of the equation. Thus updates to each state can be applied independently, selectively, and/or in parallel. Connections between this construction and fixed-point smoothing will be elucidated in Section~\ref{subsec:fixedpoint}. This decomposability is limited to the case of affine maps, however. In general, i.e., for the nonlinear dense EnTS in Equation~\ref{eq:composite_map_dense_update}, the smoothing update requires a coupled transformation of all the states, based on inversion of the map $\SKR_{\x_{1:t}}(\y_t^*,\cdot)$.

\subsection{Backward-in-time ordering}\label{sec:ordering_A} 

\begin{figure}[!ht]
  \centering
  \includegraphics[width=\textwidth]{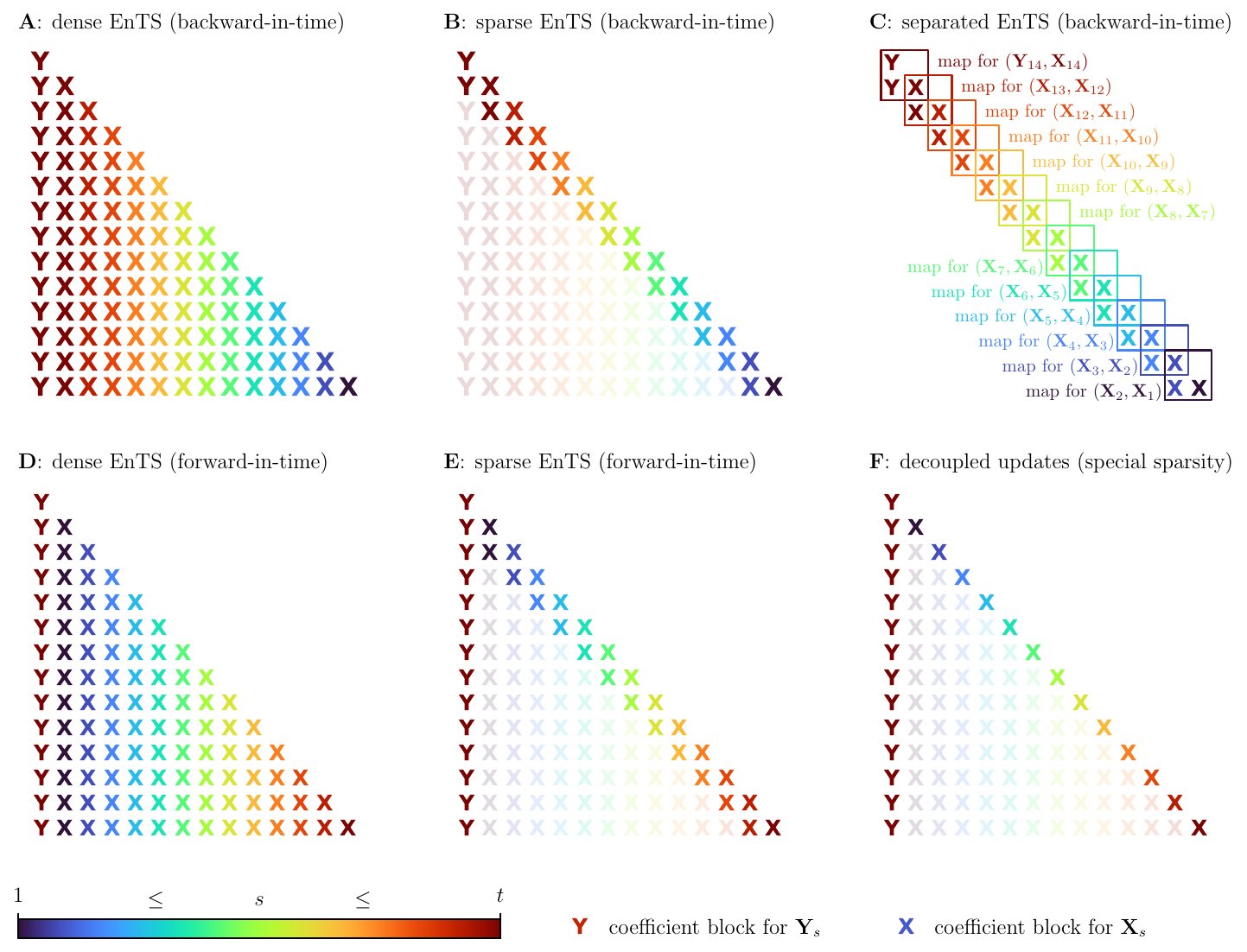}
  \caption{Different sparsity patterns in the triangular map $\SKR(\w)$, where $\w=[\y_{t},\x_{t},\dots ,\x_{1}]$ (ordering A) or $\w=[\y_{t},\x_{1},\dots ,\x_{t}]$ (ordering B). Color denotes the (temporal) ordering of the state blocks. In the linear case, these patterns correspond to sparsity in the coefficient matrix $\mathbf{C}$ used in the linear transport map $\SKR (\y_{t},\x_{1:t})=\mathbf{C}\w$.}
  \label{fig:sparsity_patterns}
\end{figure}

Now we consider specific orderings of the state. First, suppose that we arrange the state blocks in reverse-time along the graph as $[\y_{t},\x_{t},\x_{t-1},\dots ,\x_{1}]$. The joint pdf $p(\y_{t},\x_{1:t} \vert \y_{1:t-1}^{*})$ may then be factorized in a backward telescoping decomposition analogous to Equation~\ref{eq:marginal_conditional_telescopic}:
\begin{align}
\begin{split}    
p(\y_{t},\x_{1:t}|\y_{1:t-1}^{*})  = &  \ p(\y_{t}|\y_{1:t-1}^{*})p(\x_{t}|\y_{t},\y_{1:t-1}^{*})p(\x_{t-1}|\y_{t},\x_{t},\y_{1:t-1}^{*})p(\x_{t-2}|\y_{t},\x_{t-1:t},\y_{1:t-1}^{*}) \\
& \ \cdots \ p(\x_{1}|\y_{t},\x_{2:t},\y_{1:t-1}^{*}).
\end{split}
    \label{eq:backward_decomposition}
\end{align}
Consider now the following conditional independence properties encoded in the hidden Markov model of Figure~\ref{fig:Markovian_graph}: past state blocks are conditionally independent of the most recent observation given the most recent state (i.e., $\x_{1:t-1} \ci \y_{t} \, | \, \x_{t}$) and each state block 
is conditionally independent of non-neighboring state blocks given its neighbor (i.e., $\x_{1:s-1} \ci \x_{s+1:t} \, | \, \x_{s}$ for $s = 2,\dots,t-1$). Exploiting these properties, we can remove dependencies on non-neighboring states, and most dependencies on $\y_t$, from the factorization above. This yields an equivalent but sparser variant of the decomposition in \begin{equation}
    \begin{aligned}
        p(\y_{t},\x_{1:t}|\y_{1:t-1}^{*})  &= p(\y_{t}|\y_{1:t-1}^{*})p(\x_{t}|\y_{t},\y_{1:t-1}^{*})p(\x_{t-1}|\y_{t},\x_{t},\y_{1:t-1}^{*}) \cdots  p(\x_{1}|\y_{t},\x_{2:t},\y_{1:t-1}^{*}) \\
        &= p(\y_{t}|\y_{1:t-1}^{*})p(\x_{t}|\y_{t},\y_{1:t-1}^{*})p(\x_{t-1}|\x_{t},\y_{1:t-1}^{*}) \cdots  p(\x_{1}|\x_{2:t},\y_{1:t-1}^{*}) && \dagger \\
        &= p(\y_{t}|\y_{1:t-1}^{*})p(\x_{t}|\y_{t},\y_{1:t-1}^{*})p(\x_{t-1}|\x_{t},\y_{1:t-1}^{*}) \cdots  p(\x_{1}|\x_{2},\y_{1:t-1}^{*}) && \ddagger,
    \end{aligned} 
    \label{eq:backward_decomposition_partial}
\end{equation}
where we made use of $\dagger : (\x_{1:t-1} \ci \y_{t} \, | \, \x_{t})$ and $\ddagger : (\x_{1:s-1} \ci \x_{s+1:t} \, | \, \x_{s})$. As established in Section~\ref{subsec:brief_intro}, triangular transport maps can exploit conditional independence by removing arguments corresponding to superfluous dependencies from the map component functions. The decomposition in Equation~\ref{eq:backward_decomposition_partial} corresponds to the sparsity pattern in Figure~\ref{fig:sparsity_patterns}B, where the function in the row corresponding to $\x_s$, for $s = t-1, \dots, 1$, takes the form $\SKR_{\x_s}(\x_{s+1}, \x_s)$. Overall, this is the sparsity pattern of a map $\SKR $ that pushes forward $p(\y_{t},\x_{1:t}|\y_{1:t-1}^{*})$ to a standard Gaussian distribution of the same dimension with the reverse-time ordering for the states. To perform a single smoothing update, i.e., to condition the state sequence $\x_{1:t}|\y_{1:t-1}^{*}$ on a new observation $\y_t^{*}$, one could learn this map from the joint forecast ensemble---by minimizing Equation~\ref{eq:objective_function} while \textit{imposing} the sparsity pattern just described---and then embed the result in the composite map construction of Section~\ref{sec:transport_maps}. 

\subsubsection{Sequential backward updates}
\label{sec:seqbackward}

A more flexible alternative to the smoothing update proposed above, 
however, is to replace the single high-dimensional but sparse transport map of Figure~\ref{fig:sparsity_patterns}B with a sequence of overlapping, lower-dimensional maps, each corresponding to one of the terms in Equation~\ref{eq:backward_decomposition_partial}. 
In this approach, each term $p(\x_{s}|\x_{s+1},\y_{1:t-1}^{*})$ of Equation~\ref{eq:backward_decomposition_partial} will be sampled by a separate map.
This decomposition allows us to exploit additional conditional independence, namely that $\x_{s} \ci \y_{s+1:t} \, | \, \x_{s+1}$ for $s=1, \ldots, t-1$. In other words, the state $\x_{s}$ is independent of future observations $\y_{s+1:t}$ given the next state $\x_{s+1}$. Exploiting this property, we can further sparsify Equation~\ref{eq:backward_decomposition_partial} as
\small
\begin{equation}
\begin{aligned}
    p(\y_{t},\x_{1:t}|\y_{1:t-1}^{*}) &=
    p(\y_{t}|\y_{1:t-1}^{*})  
    p(\x_{t}|\y_{t},\y_{1:t-1}^{*})   
    p(\x_{t-1}|\x_{t},\y_{1:t-1}^{*})    
    p(\x_{t-2}|\x_{t-1},\y_{1:t-1}^{*})   
    \cdots 
    p(\x_{1}|\x_{2},\y_{1:t-1}^{*}) \\ 
    &= p(\y_{t}|\y_{1:t-1}^{*})
    p(\x_{t}|\y_{t},\y_{1:t-1}^{*})
    p(\x_{t-1}|\x_{t},\y_{1:t-1}^{*})
    p(\x_{t-2}|\x_{t-1},\y_{1:t-2}^{*})
    \cdots
    p(\x_{1}|\x_{2},\y_{1}^{*}).
\end{aligned}
\label{eq:backward_decomposition_complete}
\end{equation}
\normalsize
Focusing only on the smoothing distribution at time $t$, we can rewrite this factorization more compactly as:
\begin{equation}
p(\x_{1:t} \vert \y^{*}_{1:t}) = 
    p(\x_{t} \vert  \y_{1:t}^{*}) \prod_{s=1}^{t-1} p(\x_s \vert \x_{s+1}, \y^{*}_{1:s}).
    \label{eq:backward_smoothing_compact}
\end{equation}

We use this factorization to sample the smoothing distribution starting from time $t$ and proceeding backwards, such that each smoothing sample is in fact a backwards \textit{trajectory} of state, as follows. Suppose we have a filtering ensemble at time $t$, i.e., $\X_t \sim p(\x_t \vert \y_{1:t}^{*})$ and a collection of ensembles $(\X_{s+1},\X_{s}) \sim p(\x_{s},\x_{s+1}|\y_{1:s}^{*})$ for $s=t-1, \dots, 1$. The latter are easily obtained during a preceding (forward) filtering pass: $\X_{s}$ are samples from the filtering marginals $p(\x_{s}|\y_{1:s}^{*})$, and $\X_{s+1}$ represent their corresponding forecasts to time $s+1$, sampled from the dynamics. From each ensemble $(\X_{s+1},\X_{s})$, we can \textit{independently} learn the component function $\SKR_{\x_s}(\x_{s+1}, \x_s)$, with $\x_{s+1}$ always in the off-diagonal input position; see Figure~\ref{fig:sparsity_patterns}C for an illustration. 
The composite map approach to conditional sampling (Section~\ref{sec:condsamp}) uses this function to build a transformation $\T _{\x_{s+1}^{*}}(\x_{s+1},\x_s)$ pushing forward the joint distribution $p(\x_{s},\x_{s+1}|\y_{1:s}^{*})$ to the conditional distribution $p(\x_{s}|\x_{s+1}^{*}, \y_{1:s}^{*})$ for any value of $\x_{s+1}^{*}$. We then apply these transformations in sequence, beginning with a member $\x_t^i$ of the time $t$ filtering ensemble. Use the map built from $(\X_t, \X_{t-1})$ to sample $\x_{t-1}^i \sim p(\x_{t-1} \vert \x_t^i, \y_{1:t-1}^\ast)$. Then use the map built from $(\X_{t-1}, \X_{t-2})$ to sample $\x_{t-2}^i \sim p(\x_{t-2} \vert \x_{t-1}^i, \y_{1:t-2}^\ast)$. Continue the process backwards to time $t=1$, and repeat it for each member of the terminal filtering ensemble. 

The resulting algorithm corresponds to the single-pass formulation (Figure~\ref{fig:smoother_types}E) of the backwards smoother, and affords substantially reduced computational cost: rather than having to implement a smoothing update at every assimilation increment, it suffices to realize only a single backwards smoothing pass at the very end of the data assimilation interval. If intermediate updates are desired (as they might be in recursive online applications), we can derive intermediate multi-pass backwards smoothers (Figure~\ref{fig:smoother_types}F) which initiate smoothing updates at select, even irregular, intervals. An important but subtle point here is that one can also use samples $(\X_{s+1},\X_{s}) \sim p(\x_{s},\x_{s+1}|\y_{1:\tau}^{*})$ from a previous smoothing pass, i.e., for $\tau > s$, to build the relevant maps. The reason is that the conditional of interest, $p(\x_s \vert \x_{s+1}, \y^{*}_{1:\tau})$, is \textit{the same} for any $\tau \geq s$, due to the conditional independence property  $\x_{s} \ci \y_{s+1:t} \, | \, \x_{s+1}$, even if the joint distributions differ for different choices of $\tau \geq s$.
Note that with $\tau = t-1$, we recover the smoothing update that exploits the factorization in Equation~\ref{eq:backward_decomposition_partial}.

\begin{remark}[Fixed-lag smoothers] 
A common technique for both regularizing multi-pass backward smoothers and reducing their computational cost is to limit the number of updated states by introducing a lag parameter $L$. Instead of processing all states backwards to time $s = 1$, we only build the maps to sample the states $\x_{t-L:t}$. We can equivalently interpret this construction as reverting the composite maps that sample earlier states, $\x_{1:t-L-1}$, to the identity function. For stochastic processes with correlation times smaller than $L$, we expect this approach to approximate the smoothing distributions accurately; see~\cite{olsson2008sequential} for an analysis of sequential Monte Carlo approximation to fixed-lag smoothing distributions, which have neglible bias under mild ergodicity assumptions. 
Moreover, for each new observation, this fixed-lag algorithm only computes $L$ maps and updates $L$ states in the ensemble; hence, the cost of each step of recursive smoothing remains constant when increasing the time $t$.
\end{remark}

\subsubsection{Relation to the ensemble Rauch-Tung-Striebel smoother (EnRTSS)}
\label{sec:enrts}

Section \ref{sec:densetoEnKS} described how the dense EnTS reverts to the EnKS when the former is restricted to affine maps. In a similar vein, the single-pass backwards EnTS presented above corresponds to a Kalman-type smoother known as the ensemble Rauch--Tung--Striebel smoother (EnRTSS), an ensemble version of the analytic Rauch--Tung--Striebel smoother \citep{Rauch1965MaximumSystems}. Applying Equation~\ref{eq:conditioning_Gaussian_samples}, the simulation of each conditional factor in Equation~\ref{eq:backward_smoothing_compact} (or Figure~\ref{fig:smoother_types}C) can be expressed as:
\begin{equation}
\begin{aligned}
    \X_{t}^{*} & = \X_{t} - \boldsymbol{\Sigma}_{\x_{t},\y_{t}}\boldsymbol{\Sigma}_{\y_{t},\y_{t}}^{-1}(\Y_{t} - \y_{t}^{*}), &&   \\[3pt]
    \X_{s}^{*} & = \X_{s} - \boldsymbol{\Sigma}_{\x_{s},\x_{s+1}}\boldsymbol{\Sigma}_{\x_{s+1},\x_{s+1}}^{-1}(\X_{s+1} - \X_{s+1}^{*}), &&  s < t,
\end{aligned}
\label{eq:mp_EnRTSS_from_TM}
\end{equation}
where the first equation ($s=t$) realizes the EnKF, and the second equation ($s<t$) corresponds to the backwards EnRTSS update, conditioning $\X_s$ on $\X_{s+1}^\ast$. An important difference from the EnKS is that the EnRTSS update in Equation~\ref{eq:mp_EnRTSS_from_TM} relies on the cross-covariances between states at successive times, rather than the cross-covariances between states and observations at different times (see Equation~\ref{eq:firstEnKS}). This means that the EnRTSS must be applied in sequence \citep{Sarkka2010BayesianSmoothing} whereas the EnKS can be applied in parallel, but the former has important numerical advantages that we will examine in the experimental section of this manuscript.

Depending on whether the ensembles $(\X_{s+1},\X_{s})$ are drawn from a previous smoothing distribution $p(\x_{s+1},\x_{s}|\y_{1:t-1}^{*})$ (for $t>s+1$) or a forecast-augmented filtering distribution $p(\x_{s+1},\x_{s}|\y_{1:s}^{*})$, we derive the single-pass or multi-pass formulation of the EnRTSS \citep[e.g., ][]{Raanes2016OnSmoother}.\footnote{Recall that these ensembles are also used to estimate the covariance and cross-covariance terms in Equation~\ref{eq:mp_EnRTSS_from_TM}; hence these covariances are more precisely conditional covariances, e.g., $\boldsymbol{\Sigma}_{\x_s, \x_{s+1} \vert \y^{*}_{1:s}}$ or $\boldsymbol{\Sigma}_{\x_s, \x_{s+1} \vert \y^{*}_{1:t-1}}$, and so on. For simplicity, and to accommodate different single- and multi-pass variants of the algorithm, we leave this conditioning implicit in Equation~\ref{eq:mp_EnRTSS_from_TM}.} 
Assuming a linear forecast model $\x_{s+1} = \mathbf{M}\x_{s} + \boldsymbol{\epsilon}$ with additive and independent Gaussian noise $\boldsymbol{\epsilon} \sim \mathcal{N}(\mathbf{0},\mathbf{Q})$, the second part of Equation~\ref{eq:mp_EnRTSS_from_TM} can be expanded as 
\begin{equation}
    \X_{s}^{*} = \X_{s} - \boldsymbol{\Sigma}_{\x_{s},\x_{s}}\mathbf{M}^{\top }\left(\mathbf{M}\boldsymbol{\Sigma}_{\x_{s},\x_{s}}\mathbf{M}^{\top } + \mathbf{Q}\right)^{-1}\left(\mathbf{M}\X_{s} + \boldsymbol{\epsilon} - \X_{s+1}^{*}\right), \quad s < t,
    \label{eq:EnRTSS_expanded}
\end{equation}
which recovers the conventional EnRTSS formulation.

\subsection{Forward-in-time ordering}\label{sec:ordering_B}

An alternative ordering arranges the variable blocks as $[\y_{t},\x_{1},\x_{2},\dots ,\x_{t}]$ to obtain a \textit{forward} telescoping decomposition of the joint pdf $p(\y_{t},\x_{1:t} \vert \y_{1:t-1}^{*})$:
\begin{align}
\begin{split}
    p\left(\y_{t},\x_{1:t}|\y_{1:t-1}^{*}\right) = & \  
    p\left(\y_{t}|\y_{1:t-1}^{*}\right)p\left(\x_{1}|\y_{t},\y_{1:t-1}^{*}\right)p\left(\x_{2}|\y_{t},\x_{1},\y_{1:t-1}^{*}\right)p\left(\x_{3}|\y_{t},\x_{1:2},\y_{1:t-1}^{*}\right) \\ & \ \cdots \ p\left(\x_{t}|\y_{t},\x_{1:t-1},\y_{1:t-1}^{*}\right).
\end{split}
    \label{eq:joint_analysis_decomposition}
\end{align}
The sparsity pattern of the corresponding transport map is illustrated in Figure~\ref{fig:sparsity_patterns}D. Once again, we may use conditional independence, specifically the Markov property of the state variables, $\x_{s} \ci \x_{1:s-2} \, | \, \x_{s-1}$ for $s = 2,\dots, t$, to remove some variable dependencies from the pdf factorization and the transport map, as illustrated in the following equation and in Figure~\ref{fig:sparsity_patterns}E. Unlike the backward ordering, however, we \textit{cannot} remove the dependence on $\y_{t}$ from any of the marginal conditionals: 
\begin{equation}
\begin{aligned}
    p(\y_{t},\x_{1:t}|\y_{1:t-1}^{*}) &=
    p(\y_{t}|\y_{1:t-1}^{*})
    p(\x_{1}|\y_{t},\y_{1:t-1}^{*})
    \cdots
    p(\x_{s}|\y_{t},\x_{1:s-1},\y_{1:t-1}^{*})
    \cdots
    p(\x_{t}|\y_{t},\x_{1:t-1},\y_{1:t-1}^{*})
    \\
    &= p(\y_{t}|\y_{1:t-1}^{*})
    p(\x_{1}|\y_{t},\y_{1:t-1}^{*})
    \cdots
    p(\x_{s}|\y_{t},\x_{s-1},\y_{1:t-1}^{*}) 
    \cdots  
    p(\x_{t}|\y_{t},\x_{t-1},\y_{1:t-1}^{*}).
\end{aligned}
\label{eq:joint_analysis_decomposition_complete}
\end{equation} 
Focusing on the conditional that represents the smoothing distribution at time $t$, we can rewrite the factorization compactly as:
\begin{equation}
p(\x_{1:t} \vert \y^{*}_{1:t}) = p(\x_1 \vert \y^{*}_{1:t}) \prod_{s=2}^{t} p(\x_s \vert \x_{s-1}, \y^{*}_{1:t}).
    \label{eq:forward_smoothing_compact}
\end{equation}

\subsubsection{Sequential forward updates}
\label{sec:seqforward}

As before, we can employ the sparse factorization in Equation~\ref{eq:joint_analysis_decomposition_complete} to construct a map that transforms samples from $p(\y_t,\x_{1:t}|\y_{1:t-1}^*)$ into samples from the smoothing distribution $p(\x_{1:t}|\y_{1:t}^*)$. In contrast to the backwards smoother, the forward-in-time ordering will produce smoothing samples that are \textit{forward} trajectories of the state. More precisely, given a value of $\y_t^*$, a sample of the sequence $\x_{1:t}$ conditioned on the realized observations $\y_{1:t}^*$ is generated by sampling the first state $\x_1^i \sim p(\x_1|\y_{1:t}^*)$ and sequentially sampling $\x_s^i \sim p(\x_s|\y_{1:t}^*,\x_{s-1}^i)$ from each term in Equation~\ref{eq:joint_analysis_decomposition_complete}, for $s=2,\dots,t$.

To generate these smoothing samples using composite maps, we first construct a triangular map $\SKR $ that pushes forward $p(\y_t, \x_{1:t}|\y_{1:t-1}^*)$ to a standard Gaussian of the same dimension. Each component of $\SKR $ has the sparse variable dependence shown Figure~\ref{fig:sparsity_patterns}E, corresponding to Equation~\ref{eq:joint_analysis_decomposition_complete}. That is, the function in the row corresponding to $\x_{s}$ only depends on $\x_s$, the previous state $\x_{s-1}$, and the observation $\y_t$; it is $\S _{\x_s}(\y_t,\x_{s-1},\x_s)$. Each such function can be learned from ensemble members $(\X_{s-1},\X_{s},\Y_t) \sim p(\x_{s-1},\x_{s}, \y_{t}|\y_{1:t-1}^*)$, which arise from a previous smoothing pass and forecasting the observation $\y_t$ (see Section~\ref{subsec:subsequent_iterations}). Given such a map, the composite map approach in Section~\ref{sec:condsamp} yields a transformation $\T _{\y_t^*,\x_{s-1}^*}(\y_t,\x_{s-1},\x_{s})$ for sampling the states sequentially. This transformation pushes forward samples from the joint distribution $p(\x_{s-1},\x_{s},\y_t|\y_{1:t-1}^*)$ to the conditional distribution $p(\x_{s}|\x_{s-1}^*,\y_{1:t}^*)$ for any value of $\x_{s-1}^*$ and $\y_t^*$. We apply these transformations in a forward sequence, beginning with a sample $(\x_1^i, \y_t^i) \sim p(\x_1, \y_t \vert \y^{*}_{t-1})$ from the prior smoothing ensemble.
For the first state, we evaluate 
\begin{equation}
    \x_1^{*,i} = \S _{\x_1}(\y_t^*,\cdot)^{-1} \circ \S _{\x_1}(\y_t^i,\x_1^i).
\end{equation} 
Subsequent states in the $i$th forward trajectory are sampled as  
\begin{equation}
\x_{s}^{*,i} = \T _{\y_t^*,\x_{s-1}^{*,i}}(\y_t^i,\x_{s-1}^i,\x_s^i) = \S _{\x_s}(\y_t^{*},\x_{s-1}^{*,i},\cdot)^{-1} \circ \S _{\x_s}(\y_t^i,\x_{s-1}^i,\x_s^i).
\label{eq:generalFWcomposite}
\end{equation}

\begin{remark}
Each component function $\S_{\x_s}$ of the triangular map described above pulls back a standard normal reference to a conditional pdf representing the \textit{modified} transition dynamics $p(\x_s|\x_{s-1},\y_t)$, i.e., the dynamics conditioned on a value of the future observation $\y_t$. For dynamical models where the state evolves deterministically, i.e., $\x_s = f(\x_{s-1})$ for some function $f$, we simply have $p(\x_s|\x_{s-1},\y_t) = \delta(\x_{s} - f(\x_{s-1}))$, independent of $\y_t$. 
\end{remark}

\begin{remark}
As with the backwards-in-time smoother, we may also introduce a lag parameter to constrain the update to states $\x_{t-L:t}$ given an observation $\y_t$. For $L = 0$, we recover a filtering algorithm, while $L = 1$ corresponds to sampling directly from the lag-1 smoothing distribution $p(\x_{t-1}|\y_{1:t^*})$. While lag-1 smoothing can also be realized with the backward smoother, 
the forward-in-time approach does not require first sampling from the filtering distribution.
\end{remark}

\subsubsection{Special case: linear forward updates}

As for the backward-in-time ordering, we consider what happens to the general transport-based forward smoothing algorithm when we restrict the map $\S$ to be an affine function. Let the first row block be $\S _{\x_1}(\y_t, \x_1) = \A_1(\x_1 + \K_1\y_t)$ and let the subsequent rows have the affine form
\begin{equation} \label{eq:FIT_linearmap}
\S _{\x_s}(\y_t, \x_{s-1}, \x_s) = \A_s(\x_s + \K_s\y_t + \B_s\x_{s-1} + \mathbf{c}_s),\quad s = 2,\dots,t.
\end{equation}
As shown in  the following proposition (see proof in Appendix~\ref{sec:AppendixD}), the matrices $\K_s,\B_s$ obtained by minimizing the objective in Equation~\ref{eq:objective_function} over all such affine maps are given by the solution of a linear system (Equation~\ref{eq:coupledgain_FIT}) involving the joint covariance of $(\x_s,\x_{s-1},\y_t)$:
\begin{proposition} \label{prop:forwardupdates}
Let $\bfSigma$ denote the joint covariance matrix of $(\x_{s-1},\x_s,\y_t)$. The minimizer of Equation~\ref{eq:objective_function} over all affine maps of the form $\SKR_{\x_s}(\y_t,\x_{s-1},\x_s) = \A_s(\K_s\y_t + \B_s\x_{s-1} + \x_s + \mathbf{c}_s)$, with $\A_s \in \R^{D \times D}$, $\K_s \in \R^{D \times O}$, $\B_s \in \R^{D \times D}$, and $\mathbf{c}_s \in \R^{D}$, is given by the solution to the linear system
\begin{equation} \label{eq:coupledgain_FIT}
    \begin{bmatrix} \bfSigma_{\y_t,\y_t} & \bfSigma_{\y_t,\x_{s-1}} \\ \bfSigma_{\y_t,\x_{s-1}} & \bfSigma_{\x_{s-1},\x_{s-1}} \end{bmatrix} \begin{bmatrix} \K_s^\top \\ \B_s^\top \end{bmatrix} = -\begin{bmatrix} \bfSigma_{\y_t,\x_{s}} \\ \bfSigma_{\x_{s-1},\x_{s}}\end{bmatrix}.
\end{equation}
\end{proposition}]
When the resulting affine triangular map is then used to construct a composite map, $\mathbf{A}_s$ and $\mathbf{c}_s$ drop out and only the optimal $\mathbf{K}_s, \mathbf{B}_s$ remain. The resulting analog of Equation~\ref{eq:generalFWcomposite}, applied to the ensemble $(\X_{s-1},\X_{s},\Y_t)$, takes the form
\begin{align}
\X_{1}^* &= \X_1 - \bfSigma_{\x_1,\y_t}\bfSigma_{\y_t,\y_t}^{-1}(\y_t^* - \Y_t), \nonumber \\
\X_{s}^* &= \X_s - \mathbf{K}_s(\y_t^* - \Y_t) - \mathbf{B}_s(\X_{s-1}^* - \X_{s-1}), \quad s = 2,\dots,t. \label{eq:forward_update_linear}
\end{align}
This series of transformations exactly condition $\X_{1:t}$ on the observation $\y_t^{*}$ in the Gaussian case. Outside of the jointly Gaussian setting, these transformations provide an approximation to the smoothing distribution, which differs from the approximation produced by the linear backwards (i.e., EnRTS) smoother of Section~\ref{sec:enrts}. We refer to this affine update in our numerical results as the ensemble forward-analysis smoother (EnFAS).

An elegant property of the forward update is that it naturally adjusts the window of updated states. If states at the beginning of the smoothing window are uncorrelated with the observation, we have $\boldsymbol{\Sigma}_{\x_{1:L},\y_t} = \mathbf{0}$ for some $1 \leq L < t$. The resulting Kalman gains satisfy $\K_s = \mathbf{0}$ (see Appendix~\ref{sec:AppendixD}) and these states are not updated, i.e., $\X_{s}^* = \X_{s}$ for $s = 1,\dots,L$. 

\subsection{Fixed point smoothing}
\label{subsec:fixedpoint}

Another common goal in applications is to recursively characterize a specific marginal of the smoothing distribution $p(\x_{1:t}|\y_{1:t}^*)$. For example, we may be interested characterizing the state at a time $s < t$, with pdf $p(\x_s|\y_{1:t}^*)$. We now outline an approach to sample from this distribution as we collect a sequence of observations at successive times. 

\subsubsection{Transport approach}
\label{subsec:fixedpointtransport}

Consider the joint distribution of the latest observation $\y_t$ with the states $(\x_s,\x_t)$ for some $1 \leq s < t$, conditioned on $\y^*_{1:t-1}$. We can decompose this joint distribution as 
\begin{equation}
    p(\y_t,\x_s,\x_t|\y_{1:t-1}^*) = p(\y_t|\y_{1:t-1}^*)p(\x_t|\y_t,\y_{1:t-1}^*)p(\x_s|\x_t,\y_t,\y_{1:t-1}^*).
\end{equation}
From the conditional independence properties encoded by the state space model in Figure~\ref{fig:Markovian_graph} we have that $\y_t \ci \x_s \vert \x_t$. Hence, an equivalent factorization is given by 
\begin{equation}
\begin{aligned}
    p(\y_t,\x_s,\x_t|\y_{1:t-1}^*) &=
    p(\y_t|\y_{1:t-1}^*)
    p(\x_t|\y_t,\y_{1:t-1}^*)
    p(\x_s|\x_t,\y_t,\y_{1:t-1}^*) \\
    &= p(\y_t|\y_{1:t-1}^*)
    p(\x_t|\y_t,\y_{1:t-1}^*)
    p(\x_s|\x_t,\y_{1:t-1}^*).
\end{aligned}
\label{eq:sparse_fixedpoint_marginal}
\end{equation}

As above, we now construct a triangular transport map that characterizes the conditional $p(\x_s,\x_t|\y_t, \y_{1:t-1}^*)$.  Let $\S $ be a transport map that pushes forward $ p(\y_t,\x_s,\x_t|\y_{1:t-1}^*)$ to a standard Gaussian distribution of the same dimension. The lower block of this map has the form
\begin{equation}
    \SKR_{\x}(\y_t,\x_s,\x_t) = \begin{bmatrix*}[l] \SKR_{\X_t}(\y_t,\x_t) \\ \SKR_{\x_s}(\x_t,\x_s) \end{bmatrix*},
    \label{eq:fixedpoint_nonlinearmap}
\end{equation}
where we have exploited the conditional independence in Equation~
\ref{eq:sparse_fixedpoint_marginal} to define a map with sparser variable dependence; see Section~\ref{subsec:brief_intro} above. Given a realization of the observation $\y_t^*$, we then push forward a joint ensemble $(\X_s,\X_t,\Y_t)$ through the composite map $\T _{\y_t^*}(\y_t,\x_s,\x_t) = \S _{\x}(\y_t^*,\cdot)^{-1} \circ \SKR_{\x}(\y_t,\x_s,\x_t)$ to produce an ensemble from the smoothing distribution $p(\x_s,\x_t|\y_{1:t}^*)$. 

The composite map generates the smoothing ensemble in two stages. The first row block of $T_{\y_t^*}$ consists of the composite map for $\S_{\x_t}$, which can be used to sample from the marginal distribution of $\x_t$, i.e., the filtering distribution. This produces samples $\x_t^{*,i} \sim p(\x_t|\y_{1:t}^*)$. The second row block samples from the conditional distribution of $\x_s$ given a sample from the filtering distribution; each resulting sample is $\x_s^{*,i} \sim p(\x_s \vert \x_t^{*,i}, \y_{1:t-1}^*)$. The latter step propagates observation information from the last state to the marginal of $\x_s$.

\subsubsection{Relation to the ensemble Kalman smoother}

Now we illuminate a connection between fixed-point smoothers and the classical ensemble Kalman smoother (EnKS) by specializing the algorithm in Section~\ref{subsec:fixedpointtransport} to affine transformations. Let each row block of $\SKR_{\x}$ (Equation~\ref{eq:fixedpoint_nonlinearmap}) be an affine function chosen to minimize the KL divergence objective and hence Equation~\ref{eq:objective_function}. The ensemble updates from the resulting composite map are then given by
\begin{align}
    \X_t^* &= \X_t - \bfSigma_{\x_t,\y_t}\bfSigma_{\y_t,\y_t}^{-1}(\Y_t - \y_t^*) \\
    \X_s^* &= \X_s - \bfSigma_{\x_s,\x_t}\bfSigma_{\x_t,\x_t}^{-1}(\X_t - \X_t^*).
\end{align}
Combining these expressions yields a direct transformation to update the time-$s$ state block:
\begin{equation} \label{eq:FixedPoint_CombinedSmoother}
    \X_s^* = \X_s - \bfSigma_{\x_s,\x_t}\bfSigma_{\x_t,\x_t}^{-1}\bfSigma_{\x_t,\y_t}\bfSigma_{\y_t,\y_t}^{-1}(\Y_t - \y_t^*), \quad s = 1,\dots,t.
\end{equation}

If, additionally, the data are described by a linear observation model $\Y_{t} = \mathbf{H}\X_{t} + \boldsymbol{\epsilon}$, with additive Gaussian noise $\boldsymbol{\epsilon}\sim\mathcal{N}\left(\mathbf{0},\mathbf{R}\right)$ that is independent of $\X_t$, the matrix gain in Equation~\ref{eq:FixedPoint_CombinedSmoother} simplifies to the EnKS gain, $\bfSigma_{\x_s,\y_t}\bfSigma_{\y_t,\y_t}^{-1}$. 
Since $s$ is an arbitrary index, we can then apply this update to \textit{each} marginal ($s=1,\ldots, t$) to draw (approximate) samples from the block conditional $p(\x_{1:t} \vert \y_{1:t}^{*})$:
\begin{equation}
    \X_{1:t}^{*} = \X_{1:t} - \bfSigma_{\x_{1:t},\y_{t}}\bfSigma_{\y_{t},\y_{t}}^{-1}\left(\Y_{t} - \y_{t}^{*}\right),
    \label{eq:EnKS_basic}
\end{equation}
where the input ensembles $(\mathbf{Y}_t,\mathbf{X}_{1:t})$ are drawn from the joint distribution $p(\y_{t},\x_{1:t}|\y_{1:t-1}^{*})$. Equation~\ref{eq:EnKS_basic} matches the  EnKS presented in Section~\ref{sec:densetoEnKS}. As described in Section~\ref{subsec:linear_maps}, this linear transformation exactly conditions $\x_{1:t}$ on $\y^*_t$ when $(\y_t, \x_{1:t})$ are jointly Gaussian, as it is identical to applying the composite map in Equation~\ref{eq:conditioning_Gaussian_samples} to the \textit{joint} state $\x_{1:t}$. In this setting, the collection of fixed-point smoothers correctly updates the \textit{joint} distribution of states, not only each individual marginal. Outside of the Gaussian setting, of course, the EnKS is in general an approximation to the Bayesian update, both jointly and marginally. 

Note that by exploiting knowledge of $\mathbf{H}$ and $\mathbf{R}$, we can also reformulate Equation~\ref{eq:EnKS_basic} into a conventional ``semi-empirical'' form:
\begin{equation}
    \X_{1:t}^{*} = \X_{1:t} - \bfSigma_{\x_{1:t},\x_{t}}\mathbf{H}^{\top }\left(\mathbf{H}\bfSigma_{\x_{t},\x_{t}}\mathbf{H}^{\top } + \mathbf{R}\right)^{-1}\left(\mathbf{H}\X_{t} + \boldsymbol{\epsilon} - \y_{t}^{*}\right).
    \label{eq:EnKS_common}
\end{equation}

\begin{remark}
In the reformulation of the EnKS in Equation~\ref{eq:EnKS_common}, the randomly sampled observations $\Y_{t}$ on the right-hand side are obscured. Perhaps for this reason, the noise $\boldsymbol{\epsilon}$ is often interpreted as a perturbation to realized observation $\y_{t}^{*}$ in ensemble data assimilation literature. But the more general form of Equation~\ref{eq:EnKS_basic} makes clear that the perturbations are part of the prior observation ensemble, rather than perturbations to $\y_t^{*}$; each member $\y_t^i$ of the ensemble $\Y_t$ is a draw from $p(\y_t \vert \x_t^i)$, where each $\x_t^i$ is in turn a member of the ensemble $\X_{1:t} \sim p(\x_{1:t} \vert \y_{1:t-1}^{*})$, as described in the opening paragraphs of Section~\ref{sec:ents_all}. While this distinction may be of little consequence in settings where the noise $\boldsymbol{\epsilon}$ is independent of $\x_t$, it can become important if, for example, the variance of the noise depends on the state value; in this case, each perturbation $\y_t^i$ must be associated with a given $\x_t^i$, which is naturally the case when both state and observations are viewed as part of an extended forecast ensemble. A similar comment holds for the EnRTSS (Section~\ref{sec:enrts}).
For more on this interpretation, see \citet{van2020consistent}. 
\end{remark}

An important property of the linear-Gaussian EnKS update is that the operation in Equation~\ref{eq:EnKS_basic} proceeds independently for each time-block of $\x_{1:t}$. As a consequence, the EnKS update can be applied independently to states at different times, in arbitrary order, and even only selectively. This fact should hardly be surprising, given that we arrived at Equation~\ref{eq:EnKS_basic} by way of fixed-point smoothing; indeed, the decomposability of the EnKS not only enables fixed-point smoothing (Figure~\ref{fig:smoother_types}C) but also underlies the fixed-lag variant of the EnKS (Figure~\ref{fig:smoother_types}B) and permits smoothing strategies based on parallelizing, aggregating, or re-ordering the smoothing updates \citep[e.g.,][]{Ravela2007FastSmoothing,Carrassi2018DataPerspectives}. We emphasize that such decompositions are possible because the Kalman-style update to a state block $\x_{s}$ above does not depend on the cross-covariance between $\x_s$ and other state blocks $\x_{r\neq s}$.

\subsubsection{Joint smoothing via decoupled updates}

The fact that the EnKS performs marginal updates of each state $\x_s$ in a way that does not depend on other states suggests that we can also recover Equation~\ref{eq:EnKS_basic} by considering linear maps with a special ``excess'' sparsity, encoded in the ansatz
\begin{equation} \label{eq:special_sparsity_S}
\S _{\x_s}(\y_t, \x_s) = \mathbf{A}_s(\mathbf{K}_s\y_t + \x_s) + \mathbf{c},
\end{equation}
for $s = 1,\dots,t$. This ansatz corresponds to the linear maps of Equation~\ref{eq:FIT_linearmap} with the additional constraint $\mathbf{B}_s = \mathbf{0}$. It can be shown that minimizing the objective in  Equation~\ref{eq:objective_function} over affine maps of this form yields $\mathbf{K}_s = -\bfSigma_{\x_s,\y_t}\bfSigma_{\y_t,\y_t}^{-1}$; see Appendix~\ref{sec:AppendixD}. The associated composite maps then take the form $\T_{\y_t^*}(\y_t,\x_s) = \x_s + \mathbf{K}_s(\y_t - \y_t^*)$. Hence, the composite maps resulting from the special sparsity of Equation~\ref{eq:special_sparsity_S} exactly match the joint linear EnKS update in Equation~\ref{eq:EnKS_basic}.

Consequently, the composite map arising from linear maps with special sparsity correctly captures the joint smoothing distribution $p(\x_{1:t} \vert \y^{*}_{1:t})$ if $\x_{1:t},\y_t \vert \y_{1:t-1}^*$ are jointly Gaussian. More specifically, the correlations between states at different times are preserved. This is surprising, since the sparsity of Equation~\ref{eq:special_sparsity_S} imposes the conditional independence $\x_s \ci \x_{1:s-1} | \y_t$, which does not generally hold for an arbitrary multivariate Gaussian! The inconsistency can be explained as follows: sampling the states $\x_{1:t}$ by pushing forward Gaussian samples through the inverse map $\S^{-1}$ does \textit{not} preserve correlations in the joint smoothing distribution. Converting the map $\S$ with this special sparsity to a composite map, however, corrects this erroneous assumption. In particular, it enables updates derived only for \textit{marginal} (time-$s$) inference to preserve \textit{joint} (across-time) information. 

In contrast, the linear maps used by the forward or backward-in-time smoothers (e.g., Equation~\ref{eq:FIT_linearmap}) sample exactly from the multivariate Gaussian smoothing distribution by either of these methods (recall Section~\ref{sec:condsamp}): pushing independent Gaussian samples through the inverse of $\S$ \textit{or} pushing prior samples through the composite map $\T$.

\begin{remark}
In non-Gaussian settings, the special sparsity pattern in Figure~\ref{fig:sparsity_patterns}F is not guaranteed to sample correctly from the joint smoothing distribution $p(\x_{1:t} \vert \y^{*}_{1:t})$, even when a nonlinear map $\S$ with this special sparsity pattern is used to build a composite map $\T$. More specifically, the composite map operating on each marginal may not be sufficient to capture the dependence between state variables, e.g., high-order cross-moments. In this setting, the maps used in the dense, forward-in-time, or backward-in-time smoothers, which each depend on at least two state variable blocks, are the \emph{consistent} way to jointly update the states.
\end{remark}

\subsection{Iterations for subsequent data} \label{subsec:subsequent_iterations}

In online scenarios, all of smoothing algorithms we have described in Sections~\ref{sec:any_ordering}--\ref{subsec:fixedpoint} need to operate again after collecting a new observation at time $t+1$. This observation increases the length of the smoothing window and results in an extended state with $t+1$ blocks. Given samples from the smoothing distribution $p(\x_{1:t}|\y_{1:t}^*)$, we first forecast the state and observations at time $t+1$. That is, we sample the state conditioned on the final step $\x_{t+1}^i \sim p(\x_{t+1}|\x_{t}^i)$, and sample a new observation $\y_{t+1}^i \sim p(\y_{t+1}|\x_{t+1}^i)$. By augmenting the ensemble, we can then condition on $\y_{t+1}^*$ using any of the sequential backward, forward, or fixed-point methods we have just described. 

Two requirements must be met to apply these transport smoothing algorithms recursively. The first is to maintain samples from the filtering distribution, i.e., conditioned samples of the state at the end of the smoothing interval $t$. Second, the algorithms must sample \textit{jointly} from the distribution of $\x_t$ and the desired states to be updated, i.e., from $p(\x_{t-L:t-1},\x_t \vert \y_{1:t}^*)$ for the forward and backward smoothers with some lag $L$, or from $p(\x_{s},\x_t \vert \y_{1:t}^*)$ for the fixed-point smoother. This second requirement ensures that we preserve dependencies between the states that are necessary to construct the transport maps for conditioning. If these criteria are satisfied, we can safely proceed recursively: it then suffices to consider the state updates given a single ``new'' observation at the end of the smoothing interval.

\section{Numerical demonstrations} \label{sec:numerics}
Now we present numerical experiments designed to illustrate the properties and practical consequences of the smoothing patterns discussed in the preceding sections. As such, we focus our attention on relatively simple, expository systems---with the goal of extracting the essential issues and avoiding further sources of complexity. For the same reason, we also eschew standard localization and inflation strategies here. Of course, such regularizations can improve performance for large systems with small ensemble sizes, but our present goal is to illustrate the intrinsic features and pitfalls of smoothing algorithms in a ``clean'' setting unobscured by additional corrections. In a companion manuscript \citep{Ramgraber2022underUpdatesb} on nonlinear ensemble transport smoothers, we will revisit regularization in conjunction with significantly more complex models.

The Python code to reproduce the experiments and figures in this study is provided in the GitHub repository: \url{https://github.com/MaxRamgraber/Ensemble-Transport-Smoothing-Part-I}. The triangular transport toolbox we used in this study is available at \url{https://github.com/MaxRamgraber/Triangular-Transport-Toolbox}.

\subsection{Smoothing algorithms} \label{sec:algorithmlist}

Our experiments consider eight different smoothing algorithms:
\begin{enumerate}
    \item a linear-map \textbf{multi-pass backward} EnTS, and its multi-pass EnRTSS counterpart,
    \item a linear-map \textbf{single-pass backward} EnTS, and its single-pass EnRTSS counterpart,
    \item a linear-map \textbf{multi-pass forward} EnTS, and its multi-pass EnFAS counterpart,
    \item a linear-map \textbf{dense} EnTS, and its EnKS counterpart.
\end{enumerate}
While the Kalman-type smoothers and their corresponding linear-map EnTS variants are formally equivalent, as described in Section~\ref{sec:ents_all}, for the purposes of this numerical exercise we use different implementations. For each linear EnTS, we explicitly build the lower block of the triangular map $\S $ by minimizing Equation~\ref{eq:objective_function}, via numerical optimization, over the space of affine maps with the appropriate prescribed sparsity pattern; then we realize the composite map explicitly via Equation~\ref{eq:composite_map}. Of course, as explained in Section~\ref{subsec:linear_maps}, these explicit operations are entirely unnecessary in practice: we already have closed-form expressions for the optimal affine triangular maps and for the corresponding composite maps. The more circuitous implementation undertaken here serves to verify the formal equivalence of Kalman-type smoothers and transport smoothers in the special case of linear updates. Both the Kalman and linear transport updates remain efficient in our implementations; for instance, the latter requires at most 0.03 sec per update in the example of Section~\ref{sec:lorenz63}, for ensemble size $N=1000$.

Also, in the following experiments, we use a fully sample-based (empirical) formulation for both transport and Kalman-type smoothers unless otherwise specified. We note that the performance of Kalman-type smoothers can be improved through the use of semi-empirical identities in the covariance terms (e.g., Equation~\ref{eq:EnRTSS_expanded} and Equation~\ref{eq:EnKS_common}). However, these identities are restricted to cases where the errors are additive, Gaussian, and independent of the state, with known covariance matrices. To retain as much generality as possible, we will only assume the ability to sample the forecast and observation models, and will estimate all quantities involved directly from samples.

\subsection{Autoregressive model}\label{sec:AR}

In our first experiment, we compare the statistical properties of ensemble smoothers with different ensemble sizes to their analytical solution.
To this end, we consider a one-dimensional linear autoregressive dynamical system with Gaussian noise, paired with a linear (identity) observation model, again with additive Gaussian noise:
\begin{equation}
    \begin{aligned}
    x_{s} &= \alpha x_{s-1} + \epsilon, \qquad  &\epsilon & \sim \mathcal{N}\left(0, 1\right)\\
    y_{s} &= x_{s} + \nu, \qquad  &\nu & \sim \mathcal{N}\left(0, 1\right),
    \end{aligned}
    \label{eq:AR_model}
\end{equation}
and $\alpha=0.9$. The initial state is endowed with a Gaussian prior distribution $x_{1} \sim \mathcal{N}(0,\sigma^2)$ with variance equal to the forecast variance of the asymptotic filtering distribution, $\sigma^2 = \frac{1}{2}(\alpha^2 + \sqrt{4 + \alpha^4})$; see~\citep[Section 4.4]{law2015data} for a derivation. 
This choice of initial condition corresponds to running a smoother after a spin-up phase where the state is estimated using a Gaussian filter. We run smoothing experiments with ensembles of size $N=100$ and $N=1000$ drawn from this initial distribution, and simulate over $t=30$ time steps. The synthetic true state and observations are drawn from the dynamics in Equation~\ref{eq:AR_model}. This simple linear-Gaussian test case permits us to calculate the exact Gaussian solution analytically using the Kalman smoother, against which we can verify the mean and covariance estimates obtained from our various ensemble smoothing algorithms.

\subsubsection{Smoothing covariances} \label{subsec:AR_covariances}

\begin{figure}[!ht]
  \centering
  \includegraphics[width=\textwidth]{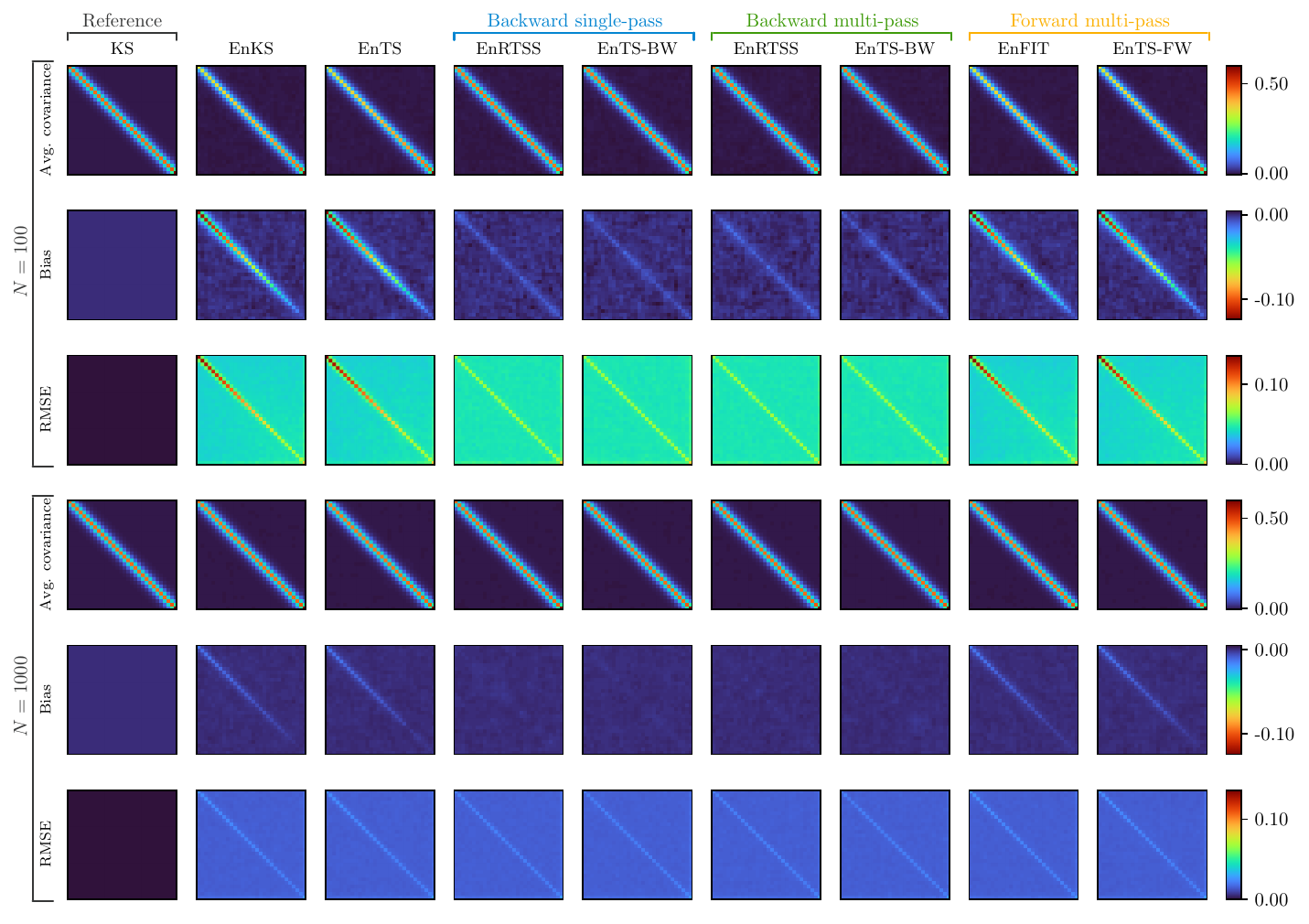}
  \caption{Estimated covariances (top row) of the autoregressive model's smoothing posterior $p(\x_{1:30}|\y_{1:30}^{*})$ averaged over $1000$ repeat simulations, along with the entrywise bias (middle row) and RMSE (bottom row) of the estimates relative to the analytical  solution for ensemble sizes $N=100$ and $N=1000$. Results are presented for the analytical reference solution and eight different smoothing algorithms: a sample-based EnKS, and its linear transport map equivalent, a sample-based single-pass and multi-pass EnRTSS, and their linear transport map equivalents, and a sample-based multi-pass forward-in-time smoother and its linear transport map equivalent.}
  \label{fig:AR_covariance}
\end{figure}

We evaluate the accuracy of our smoothing algorithms by first examining ensemble-based estimates of the covariance matrix of the smoothing distribution $p(\x_{1:30}|\y_{1:30}^{*})$. Since these estimates are random (depending on the true observations, the random initial ensemble, and realizations of the dynamical and observation noise), we run each smoothing algorithm $M=1000$ times with different random seeds. Each such simulation, with ensemble size $N$, produces a covariance estimate $\widehat{\C}^N_i$, $i=1,\ldots, M$. For each algorithm we can then evaluate the entrywise bias of the covariance estimate, $\text{bias}(\C,N) = \mathbb{E}[\widehat{\C}^N] - \C \approx \frac{1}{M} \sum_{i=1}^M \widehat{\C}_i^N - \C$, and the entrywise mean-square error of the covariance estimate, $\text{MSE}(\C,N) = \mathbb{E}[(\widehat{\C}^N - \C )^2 ] \approx  \frac{1}{M} \sum_{i=1}^M (\widehat{\C}_i^N - \C  )^2 $, where $\C$ is the true covariance computed with exact Kalman smoothing formulas (and independent of $\y_{1:30}^*$). The difference between the bias and MSE is due to fluctuation (i.e., variance) of the covariance estimates, as $\text{bias}^2 + \text{variance} = \text{MSE}$. The results are illustrated in Figure~\ref{fig:AR_covariance}, where to be concise we plot only bias and the (entrywise) square root of the MSE, i.e., the RMSE.

Analyzing the results reveals three aspects of note: 
\begin{enumerate}
    \item \textbf{Spurious correlation}: All of the sample-based covariance estimates feature a ``background'' of spurious correlations away from the diagonal of the covariance matrix. This effect is more pronounced for $N=100$ than for the $N=1000$ case (teal versus dark blue color in the RMSE plots).
    \item \textbf{Algorithmic equivalence}: Results for the linear transport smoothers and their Kalman-type counterparts are identical, verifying the theoretical equivalence between the two approaches.
    \item \textbf{Differences in robustness}: The dense and forward smoothers accumulate error; specifically, they underestimate the top-left diagonal covariance entries. These entries correspond to states at the earliest times. The finite-sample bias is more pronounced for the $N=100$ case.
\end{enumerate}

\subsubsection{Smoothing means}

\begin{figure}[!ht]
  \centering
  \includegraphics[width=\textwidth]{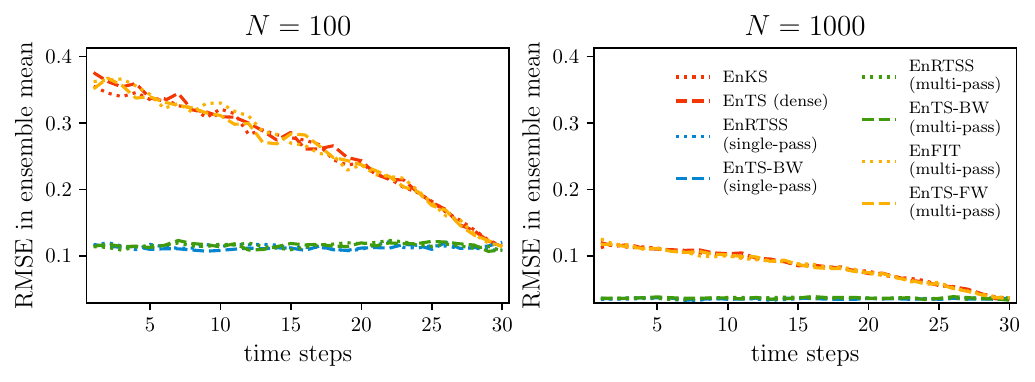}
  \caption{Entrywise RMSE for the mean estimates of the autoregressive model's smoothing posterior $p(\x_{1:30}|\y_{1:30}^{*})$ for ensemble sizes $N=100$ and $N=1000$. Results are presented for the eight smoothing algorithms in Figure~\ref{fig:AR_covariance}. \label{fig:AR_mean}}
\end{figure}

Next, we compare ensemble estimates for the mean of the smoothing distribution $p(\x_{1:30}|\y_{1:30}^*)$. For each realization of the synthetic observations $\y_{1:30}^*$, we let the ensemble mean of the states be $\widehat{\x}_i^{N}(\y_{1:30}^*)$ for $i = 1,\dots,M$. The entrywise mean-squared error (MSE) of the smoothing mean estimators is then defined as $\textrm{MSE}(N) \approx \frac{1}{M} \sum_{i=1}^{M} (\widehat{\x}_i^{N}(\y_{1:30}^*) - \mathbb{E}[\x|\y_{1:30}^*])^2$. Figure~\ref{fig:AR_mean} plots the square root of the $\textrm{MSE}$ for each time with $M = 1000$ smoothing runs. Similarly to the covariance results in subsection~\ref{subsec:AR_covariances}, we observe nearly identical performance between linear transport smoothers and their Kalman-type counterparts. As expected in a linear-Gaussian setting where all smoothers are consistent, the RMSE values decrease with increasing ensemble size $N$. Since the last smoothing marginal $\int p(\x_{1:t}|\y_{1:t}^{*}) d \x_{1:t-1}$ corresponds to the last filtering marginal $p(\x_{t}|\y_{1:t}^{*})$, all smoothers have the same RMSE at the end of the smoothing window. However, we observe the single and multi-pass backward smoothers yield lower RMSE for the states at earlier times. 
The following section will investigate the reasons for the error accumulation over time in certain smoothers.

\subsubsection{Gain and signal} \label{subsec:gain_signal}

The error accumulation observed above merits particular attention: why do we observe this effect for dense and forward smoothers, but not backward smoothers? A plausible explanation might be that this effect is a consequence of multiple, cumulative update operations: whereas dense and multi-pass smoothers have exposed the state block $\x_{s}$ to a total of $t-s+1$ update operations, single-pass backward smoothers will have only applied two update operations---once during the forward filtering pass, and once more during the backward smoothing pass. While repeated updates may indeed contribute to larger error, however, they are not solely responsible for this issue. A key counter-example is the multi-pass backward smoother, which also updates each state block $\x_{s}$ a total of $t-s+1$ times, yet does not appreciably accumulate error.

Instead, we argue that the root cause lies in how the update strategies employed by different smoothers incorporate decay in correlations. 
To investigate this, we subdivide the generic linear update (Equation~\ref{eq:conditioning_Gaussian_samples}) into two parts, a \textit{gain} term $\boldsymbol{\Sigma}_{\mathbf{b},\mathbf{a}}\boldsymbol{\Sigma}_{\mathbf{a},\mathbf{a}}^{-1}$ and a \textit{signal} term $(\mathbf{a} - \mathbf{a}^{*})$.
The gain depends on the statistical relationship between the random variables $\mathbf{a}$ and $\mathbf{b}$, and defines how the signal $(\mathbf{a} - \mathbf{a}^{*})$ affects the update to $\mathbf{b}$. To show this, we consider two scenarios under which the update operation $\mathbf{b}^{*}=\mathbf{b}-\boldsymbol{\Sigma}_{\mathbf{b},\mathbf{a}}\boldsymbol{\Sigma}_{\mathbf{a},\mathbf{a}}^{-1}(\mathbf{a} - \mathbf{a}^{*})$ will have no effect. One is the special case of zero signal, that is $\mathbf{a}=\mathbf{a}^{*}$. With the term in the parentheses being zero, we will have $\mathbf{b}=\mathbf{b}^{*}$, no matter how strong the gain. Conversely if the gain is zero because $\mathbf{a}$ and $\mathbf{b}$ are uncorrelated ($\boldsymbol{\Sigma}_{\mathbf{b},\mathbf{a}}=\mathbf{0}$), then we will likewise make no update to $\mathbf{b}$, no matter the strength of the signal. This is relevant in smoothing in which states of dynamical systems often decorrelate with time~\cite{majda2016introduction}. In particular, a smoothing update assimilating an observation $\y_{t}^{*}$ should have progressively weaker influence for states $\x_{s}$ far removed in time $s \ll t$. The backwards and dense (or forward) smoothers realize this effect quite differently, however.

To demonstrate the difference, we numerically evaluate both parts of the linear update for various ensemble smoothers. The gain is computed as the product of sample estimates of two covariance terms, $\widehat{\boldsymbol{\Sigma}}_{\mathbf{b},\mathbf{a}} \widehat{\boldsymbol{\Sigma}}_{\mathbf{a},\mathbf{a}}^{-1}$. In the present autoregressive example, this quantity is a scalar and we report its absolute deviation from zero. The signal is more subtle, as it it varies over the ensemble. Given an ensemble $\{(\a^{i}, \b^{i})\}_{i=1}^N$ and values of $\a$ upon which are conditioning, written as $\a^{*,i}$, we can capture the overall magnitude of the signal via its mean absolute deviation: $\frac{1}{N} \sum_{i=1}^N \vert \a^i - \a^{*,i} \vert$. For the EnKS at the final time, $\a^i = \y_t^i$ and $\a^{*,i} = \y_t^{*}$, independent of $i$; hence the signal is $\frac{1}{N} \sum_{i=1}^N \vert \y_t^i - \y_t^{*} \vert$, where the $\y_t^i$ are drawn from $p(\y_t \vert \y^{*}_{t-1})$. For backwards smoothers, the signal is instead related to the value of the state at the next time: $\a^{i} = \x_{s+1}$ and $\a^{*,i} = \x_{s+1}^{*,i}$, so that the signal is $\frac{1}{N} \sum_{i=1}^N \vert \x_{s+1}^i - \x_{s+1}^{*,i} \vert$. Here $\x_{s+1}^i \sim p(\x_{s+1} \vert \y^{*}_{1:\tau})$ (for some $\tau \in [s, t-1]$)\footnote{In the single-pass backward smoother, $\tau = s$; in multi-pass versions of the algorithm, $\tau > s$.} but $\x_{s+1}^{*,i} \sim p(\x_{s+1} \vert \y^{*}_{1:t})$. In other words, each $\x_{s+1}^{*,i}$ comes from the neighboring just-completed step of the backwards smoothing pass. 

\begin{figure}[!ht]
\centering
\includegraphics[width=\textwidth]{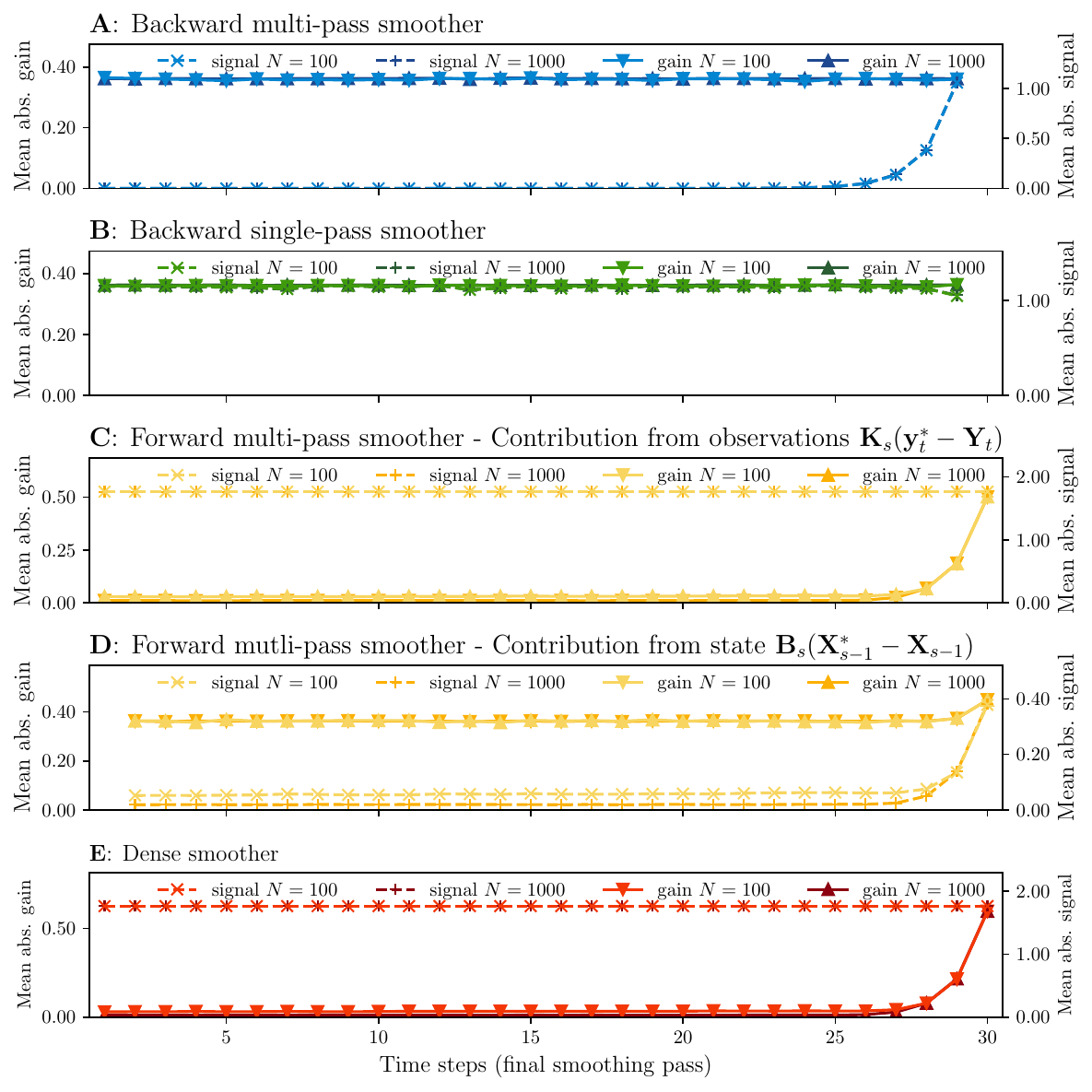}
\caption{Mean absolute deviation of the signal and gain terms for the final smoothing pass of the autoregresive model. We compare (A) the affine multi-pass and (B) single-pass backwards smoothers, (C,D) the forward smoother (for both parts of the update in Equation~\ref{eq:forward_update_linear}), and (E) the dense smoother, for ensemble sizes $N=100$ and $N=1000$.}
\label{fig:signal_vs_map_AR_results}
\end{figure}

Figure~\ref{fig:signal_vs_map_AR_results} displays the absolute deviation of the gain and signal during the final smoothing pass for ensemble sizes $N=100$ and $N=1000$. We average the results over $1000$ repeat simulations using different synthetic observations and noise variables. We summarize the differences between the two parts of the update for each ensemble smoothing algorithm below:
\begin{itemize}
    \item In \textbf{(multi-pass) backward smoothers}, the gain $\boldsymbol{\Sigma}_{\x_{s},\x_{s+1}}\boldsymbol{\Sigma}_{\x_{s+1},\x_{s+1}}^{-1}$ converges to a non-zero quantity, whose magnitude depends on the strength of correlation between successive pairs of state blocks (Figure~\ref{fig:signal_vs_map_AR_results}A). In contrast, the signal $(\x_{s+1}^{i} - \x_{s+1}^{i,*})$
    decays to zero during the backward pass for $s \ll t$ for the multi-pass smoother.
    As a result, the update does not affect the states at earlier times.
    We remark that the single-pass smoother is assimilating all observations in the backward pass, not just the latest one. Hence, the signal and the overall update for earlier time-steps is not expected to decay to zero (Figure~\ref{fig:signal_vs_map_AR_results}B). 
    \item In both \textbf{forward and dense smoothers}, the update contains the gain     $\boldsymbol{\Sigma}_{\x_{s},\y_{t}}\boldsymbol{\Sigma}_{\y_{t},\y_{t}}^{-1}$ that decays to zero for $s \ll t$ and larger ensemble sizes $N$ (Figure~\ref{fig:signal_vs_map_AR_results}C and ~\ref{fig:signal_vs_map_AR_results}E). 
    In small-sample regimes, however, the estimate of the cross-covariance $\boldsymbol{\Sigma}_{\x_{s},\y_{t}}$ and thus the gain is susceptible to spurious correlations, as observed with the convergence of the absolute gain to a small non-zero value for $N = 100$. These spurious correlations produce a statistical error in the analysis ensemble (e.g., the offset in the estimated smoothing mean and covariance seen in Figure~\ref{fig:AR_covariance} and~\ref{fig:AR_mean}) that accumulates with each observation assimilation.
\end{itemize}

As the forward smoother and dense smoother share a susceptibility to spurious correlation, we limit our investigation only to the dense and backward EnTS variants in the following experiments.

\subsection{Lorenz-63 model}
\label{sec:lorenz63}

Now we examine the different smoothing algorithms on the Lorenz-63 model, which has nonlinear dynamics and non-Gaussian states~\citep{Lorenz1963DeterministicFlow}. This dynamical system has a three-dimensional state $\x(t) = (x^a(t),x^b(t),x^c(t))$ that evolves in time according to the ODE system:
\begin{equation}
    \frac{d x^{a}}{d t} = \sigma\left(x^{b} - x^{a}\right), \quad
    \frac{d x^{b}}{d t} = x^{a}\left(\rho - x^{c}\right)-x^{b}, \quad
    \frac{d x^{c}}{d t} = x^{a}x^{b} - \beta x^{c},
    \label{eq:L63_scalar}
\end{equation}
where $(\sigma,\beta,\rho)$ are specified model parameters. In our experiments, we set $\sigma=10$, $\beta=\frac{8}{3}$, and $\rho=28$, for which the model exhibits chaotic dynamics. In our simulations, a single forecast step integrates these dynamics forward for $\Delta t = 0.1$ time units, using two steps of a fourth-order Runge-Kutta scheme with timestep $0.05$. We index the discretized state variable based on the number of applied forecast steps, i.e., $\x_s = \x(\Delta t (s-1))$ for $s \geq 1$.

The initial state $\x_1$ is drawn from a standard Gaussian prior distribution $p(\x_1)$. Following the setup of \citet{Lei2011ABickel}, we measure all elements of the state with independent observational errors following the observation model 
\begin{equation}
    \y_{s} = \x_{s} + \boldsymbol{\nu}, \quad \boldsymbol{\nu} \sim \mathcal{N}\left(\mathbf{0}, 4\mathbf{I}_3\right).
    \label{eq:L63_observation}
\end{equation}
As is common in data assimilation studies, we consider an \textit{identical twin} experiment where the sequence of true hidden states $\x_{s}^{\textrm{true}}$ and synthetic observations $\y_s^* \sim \mathcal{N}(\x_s^{\textrm{true}},4\mathbf{I}_3)$ are drawn from the same forecast and observation models that are used for smoothing~\citep{bengtsson1981dynamic,reich2015probabilistic} 

We run the models for $t = 2000$ time-steps to generate a sequence of true states and synthetic observations. We discard the first $1000$ steps as spin-up and run the smoothing algorithms on the last $1000$ steps. We report the square root of the mean-square error (RMSE) for estimating the true hidden state over $1000$ assimilation cycles using the ensemble mean.
To reduce computational demand, the dense and multi-pass backward smoothing updates were only extended to a lag length of $100$. All smoothing algorithms use filtering samples produced by an ensemble transport filter based on sparse transport maps, which is described in Appendix~\ref{sec:AppendixE}. 

\subsubsection{State estimation}

\begin{figure}[!ht]
\centering
\includegraphics[width=\textwidth]{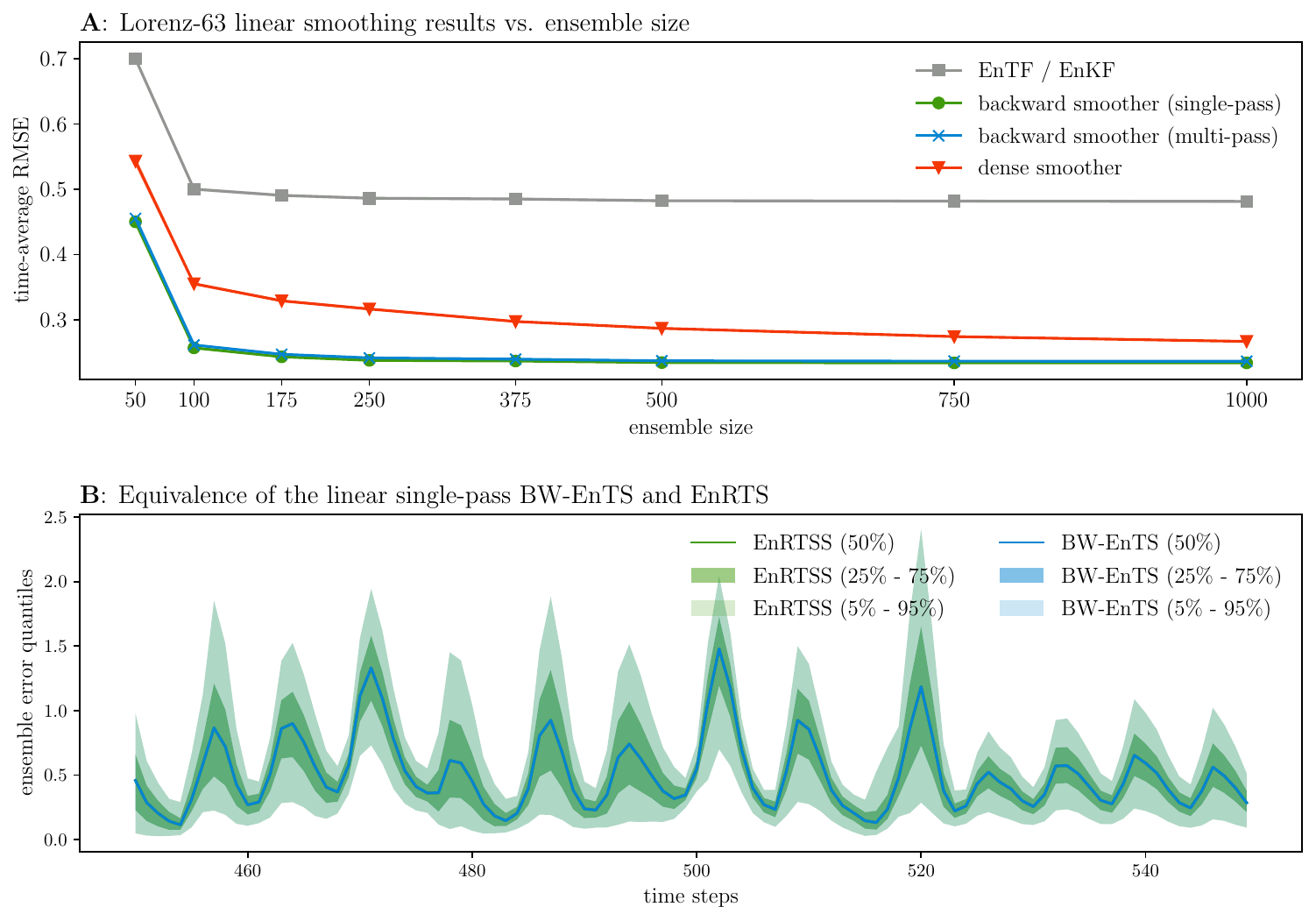}
\caption{(A) Time-averaged RMSE, and the $95\%$ confidence interval of the mean error, for different smoothers and varying ensemble sizes over $1000$ assimilation cycles. (B) Quantiles of the error between smoothing ensemble members and the true state for $N=1000$. The variability in RMSE arises from the natural oscillations in the Lorenz-63 dynamics. The RMSE values are low when the true state rests in the ``decision region'' between both wings of the attractor, and increase when the ensemble is uncertain about which wing the true state is following.}
\label{fig:results_L63}
\end{figure}

Figure~\ref{fig:results_L63} presents the average RMSE over 100 independent smoothing runs with different random seeds. The shaded region around each line shows the 95\% confidence interval for the Monte Carlo estimate of the average RMSE. We note that these intervals are very narrow and indistinguishable from the average RMSE. We compare these results to the average RMSE from filtering alone. 
Overall, smoothing results in lower error for estimating the hidden state as compared to filtering. Among the different smoothing strategies,
the single-pass and multi-pass formulations of the backward smoother provide similar performance, which suggests that error does not accumulate from multiple smoothing passes in the multi-pass formulation. Moreover, both backward smoothers outperform the dense smoother, particularly in the small ensemble size regime.

Figure~\ref{fig:results_L63}B plots \textit{quantiles} of the error $\Vert \x_s^i - \x_s^{\text{true}} \Vert$ at each time $s$ from the backwards smoothers within a representative interval of 100 timesteps after assimilating all observations. 
As discussed in Section~\ref{sec:enrts}, affine backward transport smoothers and their Kalman-type EnRTSS counterparts are equivalent; this is confirmed by the perfect overlap of the quantiles in Figure~\ref{fig:results_L63}B. The plot also shows that the estimation error (specifically, the mean of its empirical distribution over the ensemble members) fluctuates over time but remains generally bounded.

\begin{figure}[!ht]
\centering
\includegraphics[width=\textwidth]{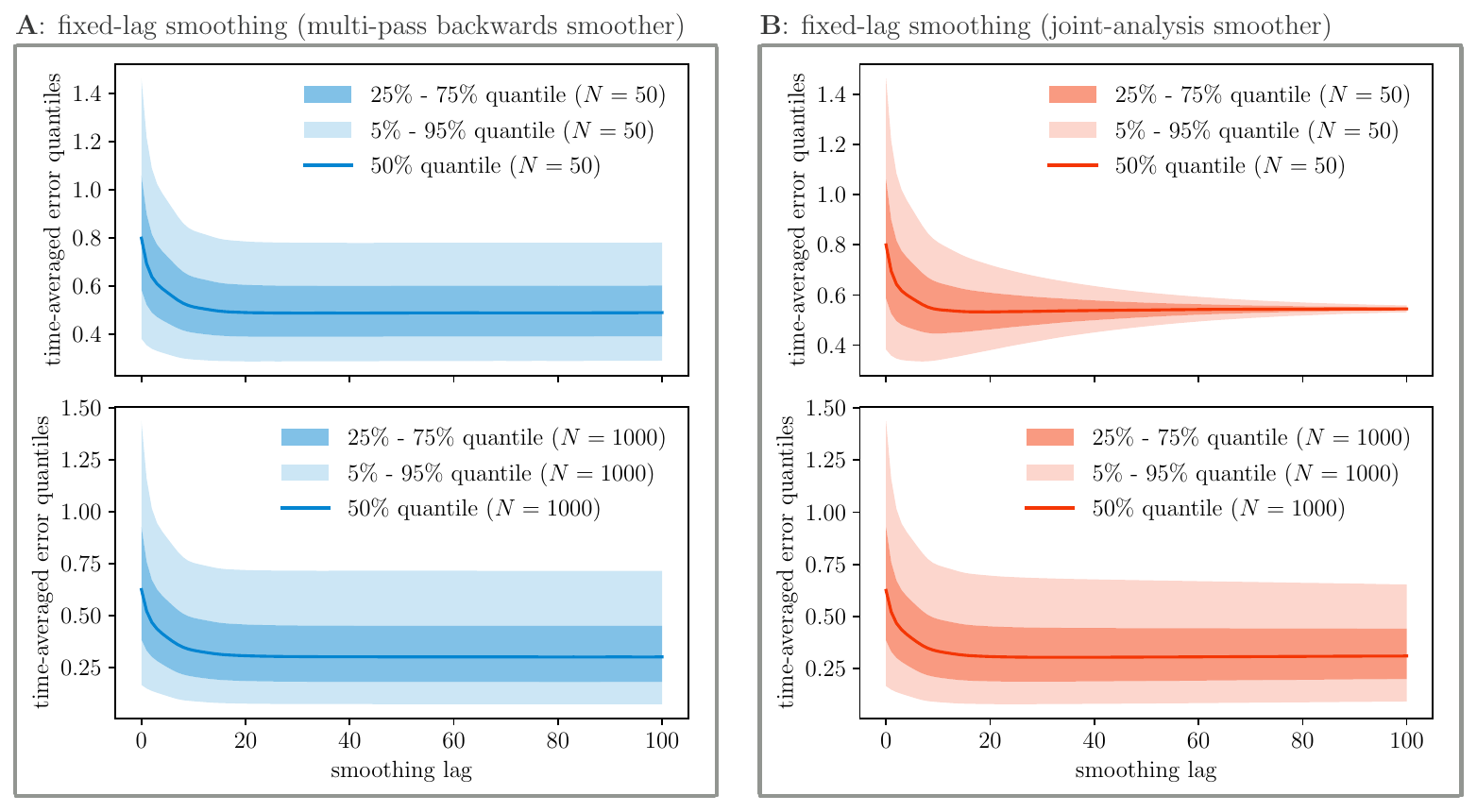}
\caption{Time-averaged quantiles of RMSE for different lag lengths and ensemble sizes for (A) the multi-pass backwards and (B) the dense smoother. As lag increases and states and predictions decorrelate, smoothing updates should attenuate and RMSE quantiles flatten out. In the backward smoother, updates become negligible with lags $L \gtrsim 15$ for both $N=50$ and $N=1000$. In the dense smoother, spurious updates cause an underestimation of uncertainty. This effect is markedly more pronounced for $N=50$ than $N=1000$.}
\label{fig:fixed_lag_L63}
\end{figure}

We now compare the performance of the fixed-lag multi-pass backward and dense smoothers for assimilating each observation. In this experiment we consider lags $L$ ranging from $0$ to $100$; we note that larger lags do not result in any further changes of the state ensemble and are unnecessary. For the smoothing pass corresponding to observation $\y_s$, we compute the $\alpha$ quantiles of the state estimation error $Q_{s,L}^\alpha \coloneqq \Vert \x_L^i - \x_L^{\text{true}} \Vert$ for lags $\max\{1,s-100\} \leq L \leq s$. Figure~\ref{fig:fixed_lag_L63} plots the average quantiles of the error over the assimilation cycles, i.e., $\overline{Q}_L^\alpha = \frac{1}{t} \sum_{s=1}^t Q_{s,L}^\alpha$ as a function of lag.
We note that the $L = 0$ point in each plot represents the filtering performance; for $L > 0$ we see the effect of smoothing. For the backward smoother, we observe that the error plateaus for larger lags and this behavior is robust to ensemble size $N$. As in the autoregressive example, this is a consequence of the decay in the backward smoother updates to zero for states far away in time from each observation.

For the dense smoother (Figure~\ref{fig:fixed_lag_L63}B), on the other hand, all of the error quantiles converge to a single value with increasing lag for $N = 50$. This suggests that the ensembles in fact collapse---perhaps as a result of spurious correlations in the estimated gains. 
We note that in this scenario, the ensemble collapse sets in before the median error plateaus. This suggests that even for an optimal choice of lag (that most reduces the RMSE), fixed-lag formulations of the dense smoother can only attenuate the collapse of the ensemble, not prevent it from happening.

\section{Discussion} \label{sec:conclusions}
Transportation of measure provides a flexible and consistent framework for Bayesian inference. Triangular transport maps, in particular, are easy to learn from samples and provide natural mechanisms for conditional sampling. This work uses such maps to construct a systematic and unified framework for Bayesian smoothing in state-space models. 

An important degree of freedom in triangular maps is the ordering of the state variables. We show that different orderings yield different smoothing approaches: an arbitrary ordering in general yields a dense nonlinear transport smoother; forward- and backward-in-time orderings of the states yield different sparse smoothing strategies that proceed sequentially through the states; and other triangular map constructions yield nonlinear fixed-lag and fixed-point smoothers. 

These smoothers are not all the same. Our numerical experiments investigate the statistical properties of various ensemble transport smoothers for two dynamical models. We show that backward smoothers are more resistant to spurious correlations than forward and dense smoothers in small ensemble size regimes---resulting in lower state estimation errors and more accurate estimates of the smoothing covariance $\boldsymbol{\Sigma}_{\x_{1:t} \vert \y^\ast_{1:t}}$. Moreover, backward smoothers obtain these results with fewer operations. While dense and multi-pass forward smoothers require $t(t+1)/2$ conditioning operations for a full smoothing pass over a fixed interval, a single-pass backward smoother only requires a forward and backward pass, adding up to $2t-1$ conditioning operations. Multi-pass backward smoothers lie somewhere in between, their cost depending on the frequency with which backward passes are initiated. 

We derived our smoothing framework for general nonlinear transport maps, which in principle can characterize the conditionals of arbitrary non-Gaussian distributions. 
Restricting the maps to be affine functions yields updates that are exact for conditional sampling in linear-Gaussian settings. We show analytically and experimentally that such affine-map ensemble transport smoothers are equivalent to established smoothing algorithms that use Kalman updates, such as the EnRTSS and the EnKS~\citep{Cosme2012SmoothingSolutions}. More general classes of transport maps, which include nonlinear functions, yield nonlinear updates and improved inference in non-Gaussian settings.

In Part II of this work, we will focus on the implementation of nonlinear updates in backward ensemble transport smoothers. While the numerical examples in the present paper deliberately focused on simple or idealized dynamical systems, in order to explore general properties of different smoothing strategies, the companion paper will demonstrate nonlinear smoothers in more complex chaotic dynamical systems. There, we will also discuss localization and regularization methods for smoothing in systems with high-dimensional states.

\section{Acknowledgements} \label{acknowledgements}

We thank the anonymous reviewers for their many thoughtful and helpful comments. The research of MR leading to these results has received funding from the Swiss National Science Foundation under the Early PostDoc Mobility grant P2NEP2 191663. RB and YM also acknowledge support from the US Department of Energy AEOLUS Mathematical Multifaceted Integrated Capabilities Center (MMICC) under award DE-SC0019303. MR and YM also acknowledge support from the Office of Naval Research Multidisciplinary University Research Initiative on Integrated Foundations of Sensing, Modeling, and Data Assimilation for Sea Ice Prediction under grant award N00014-20-1-2595.

\begin{appendices}
\section{Derivation of the objective function}\label{sec:AppendixA}

To identify the map $\SKR$ which pulls back $\eta$ to $p$, we seek to minimize the Kullback-Leibler divergence from $\SKR^{\sharp}\eta$ to $p$. That is,

\begin{equation*}
    \KLDiv(p||\SKR^{\sharp}\eta) = \mathbb{E}_{p}\left[\log\frac{p(\x)}{\SKR^{\sharp}\eta(\x)}\right].
\end{equation*}
If we split the terms in the logarithm and use the change of variables formula in Equation~1 for $\SKR^{\sharp}\eta$, we obtain
\begin{equation*}
    \KLDiv(p||\SKR^{\sharp}\eta) = \mathbb{E}_{p}\left[\log p(\x)-\log(\eta(\SKR(\x))\det {\nabla} \SKR(\x))\right].\label{eq:Kullback-Leibler1}
\end{equation*}
Since we defined $\eta$ as a standard normal distribution with log-density $\log\eta(\x) = -\frac{D}{2}\log(2\pi) -\frac{1}{2}\sum_{k=1}^{D}x_{k}^{2}$, we have
\begin{equation} \label{eq:KLobj_intermediate}
    \KLDiv(p||\SKR^{\sharp}\eta) = \mathbb{E}_{p}\left[\log p(\x) + \frac{D}{2}\log(2\pi) + \frac{1}{2}\sum_{k=1}^{D}S_{k}(\x)^{2} - \log \det {\nabla} \SKR(\x)\right].
\end{equation}
Due to the lower triangular structure of $\S$, the Jacobian ${\nabla} \S(\x)$ is a lower triangular matrix. The determinant of a lower triangular matrix is given by the product of its diagonal entries
\begin{equation*}
    \log \det {\nabla} \SKR(\x)=\log\left[\prod_{k=1}^{D}\frac{\partial S_{k}(\x)}{\partial x_{k}}\right]=\sum_{k=1}^{D}\log\frac{\partial S_{k}(\x)}{\partial x_{k}}.
\end{equation*}
Substituting this result in Equation~\ref{eq:KLobj_intermediate} and splitting the terms in the integrand that are independent of $\SKR$, we obtain:
\begin{equation} \label{eq:KL_objective}
    \KLDiv(p||\SKR^{\sharp}\eta) = \mathbb{E}_{p}\left[\log p(\x) - \log\frac{1}{(2\pi)^{D/2}} \right] +  \mathbb{E}_{p}\left[\sum_{k=1}^{D}\left(\frac{1}{2}S_{k}(\x)^{2} - \log\frac{\partial S_{k}(\x)}{\partial x_{k}} \right) \right].
\end{equation}

For the purpose of minimizing the objective in Equation~\ref{eq:KL_objective} over $\SKR$, the first expectation is a constant. By the linearity of the expectation operator, we can reverse the sum and expectation in second term as 
\begin{equation*}
    \mathbb{E}_{p}\left[\sum_{k=1}^{D}\left(\frac{1}{2}S_{k}(\x)^{2} - \log\frac{\partial S_{k}(\x)}{\partial x_{k}} \right) \right] = \sum_{k=1}^{D} \underbrace{\mathbb{E}_{p}\left[\left(\frac{1}{2}S_{k}(\x)^{2} - \log\frac{\partial S_{k}(\x)}{\partial x_{k}} \right) \right]}_{\mathcal{J}_k(S_k)}.
\end{equation*}
From this expression, we can see that the objective is a sum of $D$ objective functions $\mathcal{J}_k$ that only depend on one map component $S_k$. 
Thus, we can minimize each term $\mathcal{J}_k$ separately for one component of the transport map. Considering that we have samples from $p$, we can approximate the expectation as an arithmetic mean. This yields the empirical objective function for each map component 
\begin{equation*}
    \widehat{\mathcal{J}}_k(S_{k}) = \sum_{i=1}^{N}\left(\frac{1}{2}S_{k}(\x^{i})^{2} - \log\frac{\partial S_{k}(\x^{i})}{\partial x_{k}} \right).
\end{equation*}

\FloatBarrier
\section{Conditioning a Gaussian distribution}\label{sec:AppendixB}

Let us consider a linear transport map $\SKR(\w)=\mathbf{C}\w=\z$ which maps a multivariate Gaussian target distribution $p=\mathcal{N}(\mu,\boldsymbol{\Sigma})$ to a standard multivariate Gaussian reference distribution $\eta=\mathcal{N}(\mathbf{0},\mathbf{\mathbf{I}})$. Without loss of generality, we can assume $p$ is centered at zero (i.e., $\mu=\mathbf{0}$), as we can apply this entire operation after standardizing the random variables. In this case, the coefficient matrix $\mathbf{C}$ can be found as $\mathbf{C}=\mathbf{L}^{-1}$ and $\boldsymbol{\Sigma}=\mathbf{L}\mathbf{L}^\intercal$; see Equation~10. If we subdivide $\w$ into two blocks as $\w=[\y,\x]^{\intercal}$, we define the corresponding covariance matrix as:
\begin{equation}
    \boldsymbol{\Sigma} = \begin{bmatrix*}[l]
    \boldsymbol{\Sigma}_{\y,\y} &&
    \boldsymbol{\Sigma}_{\y,\x} \\
    \boldsymbol{\Sigma}_{\x,\y} &&
    \boldsymbol{\Sigma}_{\x,\x}
    \end{bmatrix*}.
\end{equation}
The block Cholesky decomposition of $\boldsymbol{\Sigma}=\mathbf{L}\mathbf{L}^\intercal$ is given by:
\begin{equation}
    \mathbf{L} = \begin{bmatrix*}[l]
    \boldsymbol{\Sigma}_{\y,\y}^{\frac{1}{2}} &&
    \mathbf{0} \\
    \boldsymbol{\Sigma}_{\x,\y}\boldsymbol{\Sigma}_{\y,\y}^{-\frac{1}{2}} &&
    \left(\boldsymbol{\Sigma}_{\x,\x} - \boldsymbol{\Sigma}_{\x,\y}\boldsymbol{\Sigma}_{\y,\y}^{-1}\boldsymbol{\Sigma}_{\y,\x}\right)^{\frac{1}{2}}
    \end{bmatrix*}.
    \label{apeq:block_cholesky_decomposition}
\end{equation}
The inverse of a block lower-triangular matrix is defined as
\begin{equation}
    \mathbf{L}^{-1} = 
    \begin{bmatrix*}
    \mathbf{L}_{1,1} &&  \mathbf{0} \\
    \mathbf{L}_{2,1} &&  \mathbf{L}_{2,2}
    \end{bmatrix*}^{-1} = 
    \begin{bmatrix*}
    \mathbf{L}_{1,1}^{-1} &&  \mathbf{0} \\
    -\mathbf{L}_{2,2}^{-1}\mathbf{L}_{2,1}\mathbf{L}_{1,1}^{-1} && \mathbf{L}_{2,2}^{-1}
    \end{bmatrix*}.
    \label{apeq:block_matrix_inverse}
\end{equation}
Plugging the entries of Equation~\ref{apeq:block_cholesky_decomposition} into Equation~\ref{apeq:block_matrix_inverse}, we obtain the following expression for the matrix $\mathbf{C}$:
\begin{equation}
    \begin{aligned}
    \mathbf{C} &= 
    \begin{bmatrix*}
    \mathbf{C}_{1,1} &&  \mathbf{0} \\
    \mathbf{C}_{2,1} && \mathbf{C}_{2,2}
    \end{bmatrix*} = 
    \begin{bmatrix*}
    \boldsymbol{\Sigma}_{\y,\y}^{-\frac{1}{2}} &&  \mathbf{0} \\
    -\left(\boldsymbol{\Sigma}_{\x,\x} - \boldsymbol{\Sigma}_{\x,\y}\boldsymbol{\Sigma}_{\y,\y}^{-1}\boldsymbol{\Sigma}_{\y,\x}\right)^{-\frac{1}{2}}\boldsymbol{\Sigma}_{\x,\y}\boldsymbol{\Sigma}_{\y,\y}^{-1} && \left(\boldsymbol{\Sigma}_{\x,\x} - \boldsymbol{\Sigma}_{\x,\y}\boldsymbol{\Sigma}_{\y,\y}^{-1}\boldsymbol{\Sigma}_{\y,\x}\right)^{-\frac{1}{2}}
    \end{bmatrix*}.
    \end{aligned}
    \label{apeq:coefficient_matrix_from_covariance}
\end{equation}

Next, let us construct the lower component block of the composite map. To do so, let us begin by defining
\begin{equation}
    \begin{bmatrix*}[l]
    \z_{\y} \\
    \z_{\x}
    \end{bmatrix*} = \begin{bmatrix*}[l]
    \SKR_{\y}\left(\y\right) \\
    \SKR_{\x}\left(\y,\x\right)
    \end{bmatrix*} = \begin{bmatrix*}
    \mathbf{C}_{1,1} &&  \mathbf{0} \\
    \mathbf{C}_{2,1} && \mathbf{C}_{2,2}
    \end{bmatrix*}\begin{bmatrix*}[l]
    \y \\
    \x
    \end{bmatrix*}.
\end{equation}
Constructing the composite map in Equation~8 from the second map component $\SKR_{\x}$ given a realization of $\y = \y^*$ yields
\begin{equation}
    \x^{*} = \mathbf{C}_{2,2}^{-1}\left(\mathbf{C}_{2,1} \y + \mathbf{C}_{2,2} \x - \mathbf{C}_{2,1} \y^{*}\right) = 
    \x + \mathbf{C}_{2,2}^{-1}\mathbf{C}_{2,1}\left(\y - \y^{*}\right).
\end{equation}
Finally, substituting in the coefficient matrix blocks from Equation~\ref{apeq:coefficient_matrix_from_covariance} retrieves the expression for sampling the conditional $\x|\y=\y^*$ of a multivariate Gaussian distribution:
\begin{align}
    \x^{*} &= \x - \left(\boldsymbol{\Sigma}_{\x,\x} - \boldsymbol{\Sigma}_{\x,\y}\boldsymbol{\Sigma}_{\y,\y}^{-1}\boldsymbol{\Sigma}_{\y,\x}\right)^{\frac{1}{2}}\left(\boldsymbol{\Sigma}_{\x,\x} - \boldsymbol{\Sigma}_{\x,\y}\boldsymbol{\Sigma}_{\y,\y}^{-1}\boldsymbol{\Sigma}_{\y,\x}\right)^{-\frac{1}{2}}\boldsymbol{\Sigma}_{\x,\y}\boldsymbol{\Sigma}_{\y,\y}^{-1}\left(\y - \y^{*}\right) \nonumber \\
    &= \x - \boldsymbol{\Sigma}_{\x,\y}\boldsymbol{\Sigma}_{\y,\y}^{-1}\left(\y - \y^{*}\right). \nonumber
\end{align}

\pagebreak

\section{Pseudo-code for the ensemble transport smoothers} \label{sec:AppendixC} 

\begin{algorithm}[!ht]
\SetAlgoLined
\DontPrintSemicolon
 \vspace{2 pt}
 
 \textbf{Input}: $N$ prior samples $\X_1\sim p(\x_1)$, the forecast model $p(\x_{s}|\x_{s-1})$, the observation model $p(\y_{s}|\x_{s})$, the smoothing window length $t$, and a stream of observations $\y_{1:t}^{*}$

 \For{$s = 1:t$}{
   
    \If{$ s \neq 1$} {
        \textit{Forecast state}: Sample $\x_{s}^{i} \sim p(\x_{s}|\x_{s-1}^{*,i}), \; i=1,\dots,N$\;
    }
    
    \textit{Forecast observation}: Sample $\y_{s}^{i} \sim p(\y_{s}|\x_{s}^{i}),  \; i=1,\dots,N$\;
    
    \textit{Filtering step}:\;
    a) Build and optimize the map component block $\SKR_{\x_s}(\y_{s},\x_{s})$  using $(\Y_s,\X_s)$\;
    b) Push forward $\Z_{s}=\SKR_{\x_s}(\Y_{s},\X_{s})$\;
    c) Pull back $\X_{s}^{*}=\SKR_{\x_s}^{-1}(\y_{s}^{*},\Z_{s})$\;
    
    }
    
\textit{Update samples}: Set $\X_{1:s} = \X_{1:s}^*$\;
    
 \For{$r = t-1:1$}{
   
    \textit{Backward smoothing step}:\;
    a) Build and optimize the map component block $\SKR_{\x_r}(\x_{r+1},\x_{r})$ using $(\X_{r+1},\X_r)$\;
    b) Push forward $\Z_{r}=\SKR_{\x_r}(\X_{r+1},\X_{r})$\;
    c) Pull back $\X_{r}^{*}=\SKR_{\x_r}^{-1}(\X_{r+1}^{*},\Z_{r})$\;
    
    }
 
    \caption{Single-pass backward EnTS} \label{alg:backward_smoother_single_pass}

\end{algorithm}

\begin{algorithm}[!ht]
\SetAlgoLined
\DontPrintSemicolon

\vspace{2 pt}
\textbf{Input}: $N$ prior samples $\X_1\sim p(\x_1)$, the forecast model $p(\x_{s}|\x_{s-1})$, the observation model $p(\y_{s}|\x_{s})$, the smoothing window length $t$, a stream of observations $\y_{1:t}^{*}$, and optional lag parameter $L \in \mathbb{N}^{+}$, else $L = t$.

\For{$s = 1:t$}{

\If{$ s \neq 1$} {
    \textit{Forecast state}: Sample $\x_{s}^{i} \sim p(\x_{s}|\x_{s-1}^{*,i}), \; i=1,\dots,N$\;
}

\textit{Forecast observation}: Sample $\y_{s}^{i} \sim p(\y_{s}|\x_{s}^{i}), \; i=1,\dots,N$\;

\textit{Filtering step}:\;
a) Build and optimize the map component block $\SKR_{\x_s}(\y_{s},\x_{s})$ using $(\Y_s,\X_s)$\;
b) Push forward $\Z_{s}=\SKR_{\x_s}(\Y_{s},\X_{s})$\;
c) Pull back $\X_{s}^{*}=\SKR_{\x_s}^{-1}(\y_{s}^{*},\Z_{s})$\;
    
    \For{$r = (s-1):\max \left(1, s-L\right)$}{
        \textit{Backward smoothing step}:\;
        a) Build and optimize the map component block $\SKR_{\x_r}(\x_{r+1},\x_r)$ given $(\X_{r+1},\X_{r})$\;
        b) Push forward $\Z_{r}=\SKR_{\x_r}(\X_{r+1},\X_{r})$\;
        c) Pull back $\X_{r}^{*}=\SKR_{\x_r}^{-1}(\X_{r+1}^{*},\Z_{r})$\;
    }
    
    \textit{Update samples}: Set $\X_{1:s} = \X_{1:s}^*$\;

}

\caption{Multi-pass backward EnTS} \label{alg:backward_smoother_multi_pass}
    
\end{algorithm}

\begin{algorithm}[!ht]
\SetAlgoLined
\DontPrintSemicolon

\vspace{2 pt}
\textbf{Input}: $N$ prior samples $\X_1\sim p(\x_1)$, the forecast model $p(\x_{s}|\x_{s-1})$, the observation model $p(\y_{s}|\x_{s})$, the smoothing window length $t$, a stream of observations $\y_{1:t}^{*}$, and optional lag parameter $L \in \mathbb{N}^{+}$, else $L = t$.
 
\For{$s = 1:t$}{
   
    \If{$ s \neq 1$} {
    
        \textit{Forecast state}: Sample $\x_{s}^{i} \sim p(\x_{s}|\x_{s-1}^{*,i}), \; i=1,\dots,N$\;
    }
    
    \textit{Forecast observation}: Sample $\y_{s}^{i} \sim p(\y_{s}|\x_{s}^{i}), \; i=1,\dots,N$\;
    
    \textit{Set first index}: $s_0 = \max\left(1,s-L\right)$\;
    
    \textit{Update first marginal}:\;
    a) Build and optimize the map component block $\SKR_{\x_{s_0}}(\y_s,\x_{s_0})$ using $(\Y_s,\X_{s_0})$\;
    b) Push forward $\Z_{s_0}=\SKR_{\x_{s_0}}(\Y_s,\X_{s_0})$\;
    c) Pull back $\X_{s_0}^{*}=\SKR_{\x_{s_0}}^{-1}(\y_{s}^*,\X_{s_0})$\;
    
    \For{$r = s_0+1:s$}{
        \textit{Forward smoothing step}:\;
        a) Build and optimize the map component block $\SKR_{
        \x_{r}}(\y_t,\x_{r-1},\x_{r})$ using $(\Y_t,\X_{r-1},\X_r)$\;
        b) Push forward $\Z_{r}=\SKR_{\x_r}(\Y_t,\X_{r-1},\X_{r})$\;
        c) Pull back $\X_{r}^{*}=\SKR_{x}^{-1}(\y_t^*,\X_{r+1}^{*},\Z_{r})$\;
    }
    
    \textit{Update samples}: Set $\X_{1:s} = \X_{1:s}^*$\;

}
\caption{Multi-pass forward EnTS \label{alg:forward_smoother_multi_pass}}
\end{algorithm}

\begin{algorithm}[!ht]
\SetAlgoLined
\DontPrintSemicolon

 \vspace{2 pt}
 \textbf{Input}: $N$ prior samples $\X_1\sim p(\x)$, the forecast model $p(\x_{s}|\x_{s-1})$, the observation model $p(\y_{s}|\x_{s})$, the length of smoothing interval $t$, the stream of observations $\y_{1:t}^{*}$, state index to be updated $j$
 
 \For{$s = 1:t$}{
   
    \If{$ s \neq 1$} {
        \textit{Forecast state}: Sample $\x_{s}^{i} \sim p(\x_{s}|\x_{s-1}^{*,i}), \; i=1,\dots,N$\;
    }
    \textit{Forecast observation}: Sample $\y_{s}^{i} \sim p(\y_{s}|\x_{s}^{i}), \; i=1,\dots,N$\;
    
    \textit{Smoothing step:}\;
    a) Build and optimize the map component block $\SKR_{\x}(\Y_{s},\X_{j},\X_{s})$\;
    b) Push forward $(\Z_{j},\Z_s)=\SKR_{\x}(\Y_{s},\X_{j},\X_{s})$\;
    c) Pull back $(\X_{j}^*,\X_{s}^{*})=\SKR_{\x}^{-1}(\y_{s}^{*},\Z_{j},\Z_{s})$\;
    
    }
 
    \caption{Fixed-point EnTS} \label{alg:joint-analysis_smoother}
    
\end{algorithm}

\FloatBarrier
\section{Proof of Proposition~1}\label{sec:AppendixD}

\begin{proof}
Let $\eta$ be a standard normal distribution of dimension $D$. The optimization problem for the row block corresponding to $\x_s$ is 
$$\min_{\SKR_{\x_s}} \mathbb{E}_{\x_{1:s-1},\y_t}[\KLDiv(p(\x_s|\x_{1:s-1},\y_t)||\SKR_{\x_s}^\sharp\eta)].$$
Retaining only the terms in the objective depending on $\SKR_{\x_s}$ and using the form of $\eta$, this optimization problem becomes
\begin{equation} \label{eq:linear_map_opt}
\min_{\K_s,\A_s,\B_s,\mathbf{c}_s} \mathbb{E}_{\x_{1:s},\y_t}\left[\frac{1}{2}(\K_s\y_t + \B_s\x_{s-1} + \x_s + \mathbf{c}_s)^\top \A_s^\top \A_s(\K_s \y_t + \B_s\x_{s-1} + \x_s + \mathbf{c}_s) - \log|\A_s| \right].
\end{equation}
For each setting of $\A_s,\K_s,\B_s$, the vector $\mathbf{c}$ that minimizes Equation is~\eqref{eq:linear_map_opt} is $\mathbf{c}_s = \K_s\mathbb{E}[\y_t] + \B_s\mathbb{E}[\x_{s-1}] + \mathbb{E}[\x_s]$. Without loss of generality we can take $(\x_{s-1},\x_s,\y_t)$ to have mean zero so that $\mathbf{c}_s = \0$. 

Then, the optimization problem for matrices $\K_s$ and $\B_s$ for each $\A_s$ is given by
\begin{equation} \label{eq:objective_matrixKB}
\min_{\K_s,\B_s} \mathbb{E}_{\x_s,\x_{s-1},\y_t} \, \text{trace}\left[\A_s^\top\A_s(\K_s\y_t + \B_s\x_{s-1} + \x_s)(\K_s\y_t + \B_s\x_{s-1} + \x_s)^\top\right],
\end{equation}
where we have used the cyclic property of the trace to express the (scalar) first term in Equation~\ref{eq:linear_map_opt}.
From the gradients of Equation~\ref{eq:objective_matrixKB}, we have that the optimal values of $\K,\B$ must satisfy the two coupled equations:
\begin{align} \label{eq:gradients_matrixBK}
\text{trace}\left[(\A_s^\top\A_s)(\bfSigma_{\x_s,\y_{t}} + \K_s\bfSigma_{\y_t,\y_t} + \B_s\Sigma_{\y_t,\x_{s-1}})\right] &= \0 \\
\text{trace}\left[(\A_s^\top\A_s)(\bfSigma_{\x_s,\x_{s-1}} + \K_s\bfSigma_{\y_t,\x_{s-1}} + \B_s\bfSigma_{\x_{s-1},\x_{s-1}})\right] &= \0.
\end{align}
If these hold for all invertible matrices $\A$, the remaining matrix equations inside each trace in Equation~\ref{eq:gradients_matrixBK} must be zero element-wise. Collecting the terms, we arrive at the linear system for $\K_s$ and $\B_s$ in Equation~26.
\end{proof}

A consequence of the proposition above is that for $\B_s = \0$, i.e., row blocks with ``special'' sparsity of the form $S_{\x_s}(\y_t,\x_{s-1},\x_s) = \A_s(\K_s\y_t + \x_s + \bfc_s)$, the linear system in Equation~\ref{eq:linear_map_opt} simplifies to $\bfSigma_{\y_t,\y_t}\K_s^\top = \Sigma_{\y_t,\x_s}$. Thus, in this setting we have $\K_s = -\bfSigma_{\x_s,\y_t}\bfSigma_{y_t}^{-1}$.

\FloatBarrier
\section{Sparse filtering updates from local observations}\label{sec:AppendixE}

In addition to conditional independence between blocks of states and observations at difference times (cf.\ Figure~1), triangular transport maps can exploit conditional independence 
among the state elements and observations at a given observation time.
These independencies yield additional sparsity in the map's components, 
which can further regularize the map estimation problem. 
In this section, we show how to exploit this structure for filtering the Lorenz-63 system with local observations, i.e., where each observation only depends on one element of the state.

\begin{figure}[!ht]
  \centering
  \includegraphics[width=\textwidth]{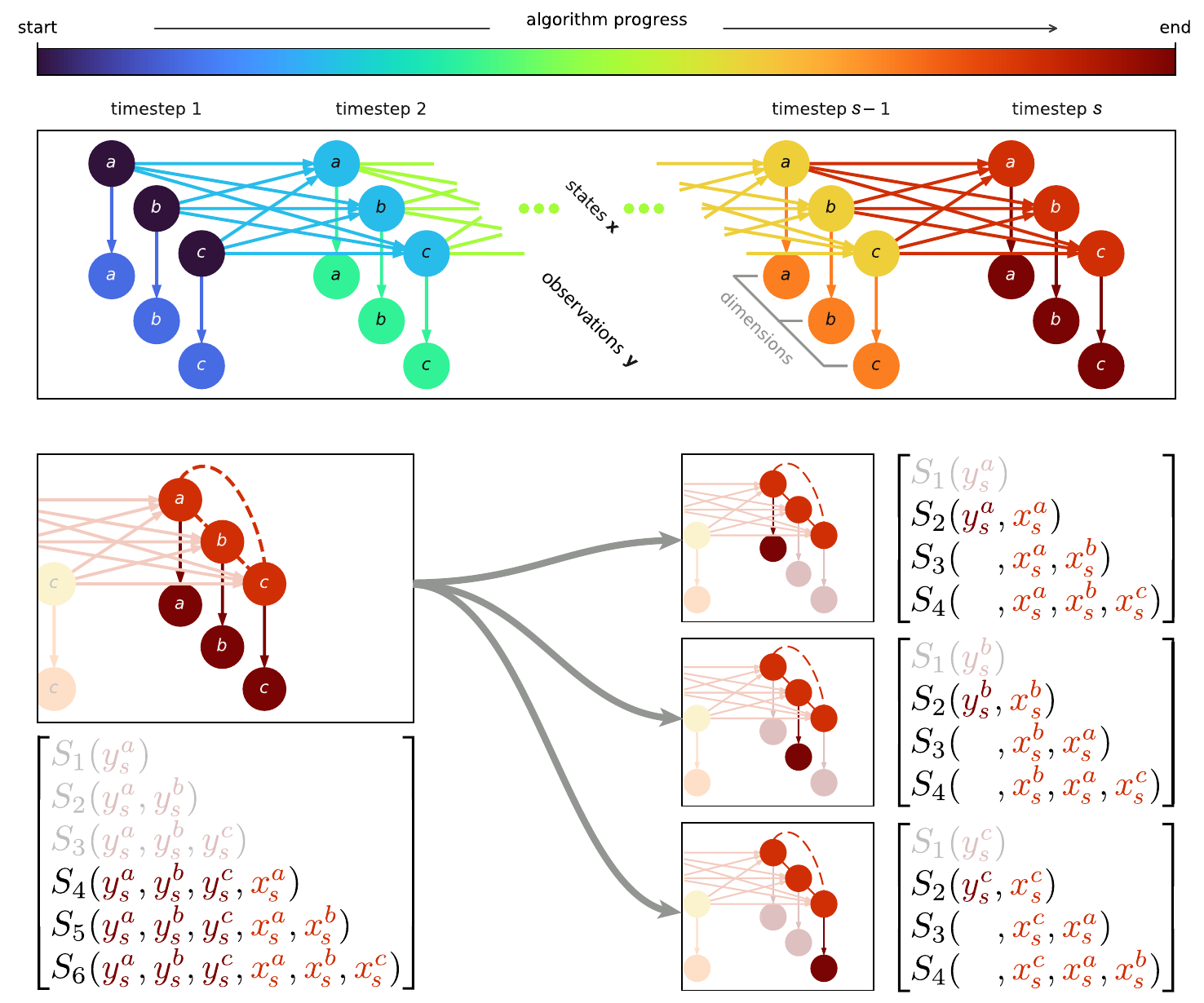}
  \caption{(Top) Probabilistic graphical model encoding the conditional dependencies between the states and observations of the Lorenz-63 system. With independent observation errors, each observation $y_{s}^{k}$, $k \in \{a, b, c \}$, is conditionally independent of $\x_s^{\neq k}$ given $x_s^k$. Assimilating observations sequentially exploits this conditional independence. Lower plots represent the undirected graphical model for the state and observations at time $t$. The single update is decomposed into $3$ updates that assimilate each scalar observation using simpler maps.}
  \label{fig:graph_Lorenz}
\end{figure}

Let the observation $\y_s = (y_s^a,y_s^b,y_s^c)$ at time $s$ be drawn from the likelihood model in Equation~39. With direct observations of the state and independent observation errors, each element of the observation $\y_s$ given the observed state is conditionally independent of the remaining states, e.g., $y_{t}^{b} \ci (x_{t}^{a}, x_{t}^{c}) \, \vert \, x_{t}^{b}$ for variable $b$. This conditional independence yields a likelihood model that factorizes as $p(\y_s|\x_s) = \prod_{k \in {a,b,c}} p(y_s^k|x_s^k)$, and enables dividing a single filtering update into three separate operations that condition the state $\x_s$ on one element of the observation at a time; see the bottom of Figure~\ref{fig:graph_Lorenz}.

Processing observations incrementally has two computational advantages. First, instead of defining a single composite map $\T_{\y_s^*}$ using Equation~8 that depends on the full observation vector, we define $\text{dim}(\y_s)$ composite maps $T_{y_s^{k,*}}$ that each have lower-dimensional inputs. For linear updates as in Equation~11, lower-dimensional maps do not require inverting the full covariance matrix for the observation. Second, processing observations serially enables us to more easily take advantage of sparsity arising from conditional independence.
As an example, consider the transport map designed to condition the full state on $y_{s}^{b,*}$.
With the variable ordering given by the observed state first followed by the remaining states (in any order), the joint pdf factorizes as   
$p(y_s^b, x_s^b, x_s^a, x_s^c) = p(y_s^b) p(x_s^b \vert y_s^b) p(x_s^a \vert x_s^b) p(x_s^c \vert x_s^b, x_s^a)$. 
This factorization guarantees that a lower triangular map of the form
\begin{equation}
    \SKR(y_{s}^{b},x_{s}^{b},x_{s}^{a},x_{s}^{c}) = 
    \begin{bmatrix*}[l]
    S_1(y_s^b) \\
    S_2(y_s^b, x_s^b) \\
    S_3( \, \ \ \, \, , x_s^b, x_s^a) \\
    S_4( \, \ \ \, \, , x_s^b, x_s^a, x_s^c)
    \end{bmatrix*},
    \label{eq:sparse_filtering_map}
\end{equation}
will exactly represent the joint distribution.
If we further make a linear ansatz for $\SKR$, we simply have the sparse lower-triangular coefficient matrix in the map:
\begin{equation}
    \SKR(y_{s}^{b},x_{s}^{b},x_{s}^{a},x_{s}^{c}) = 
    \begin{bmatrix*}
    c_{1,1} \\
    c_{2,1}     && c_{2,2} \\
                && c_{3,2}  && c_{3,3}  \\
                && c_{4,2}  && c_{4,3}  && c_{4,4}
    \end{bmatrix*}    
    \begin{bmatrix*}[l]
    y_{s}^{b} \\
    x_{s}^{b} \\
    x_{s}^{a} \\
    x_{s}^{c}
    \end{bmatrix*}.
\end{equation}

Similar reasoning gives analogous sparsity for the four-dimensional linear maps associated with the two remaining measurements, $y_{s}^{a,*}$ and $y_{s}^{c,*}$; see Figure~\ref{fig:graph_Lorenz}. In practice, the three conditioning operations at time $s$ are applied sequentially: each ``partial'' update conditions all of the state components $x_{s}^{i}$ on one component of the measurement vector $\y_{s}$. We use the updated state ensemble to draw predictive samples of the observations for the next partial update, and build the corresponding four-dimensional triangular map in Equation~\ref{eq:sparse_filtering_map}; then we apply the associated composite map to realize conditioning.
\footnote{We note that an equivalent decomposed operation is also possible with Kalman-type updates: the first step of a partial update would condition the joint distribution $p(y^{b},x^{b})$ on $y^{b,*}$ to obtain $x^{b,*}\sim p(x^{b}|y^{b,*})$. The second step then conditions $p(x^{b},x^{a},x^{c})$ on $x^{b,*}$ to obtain $x^{a,*},x^{c,*}\sim p(x^{a},x^{c}|x^{b,*}, y^{b,*})$. Then repeat this two-step process for $y^{a,*}$ and $y^{c,*}$.}

\begin{figure}[!ht]
  \centering
  \includegraphics[width=\textwidth]{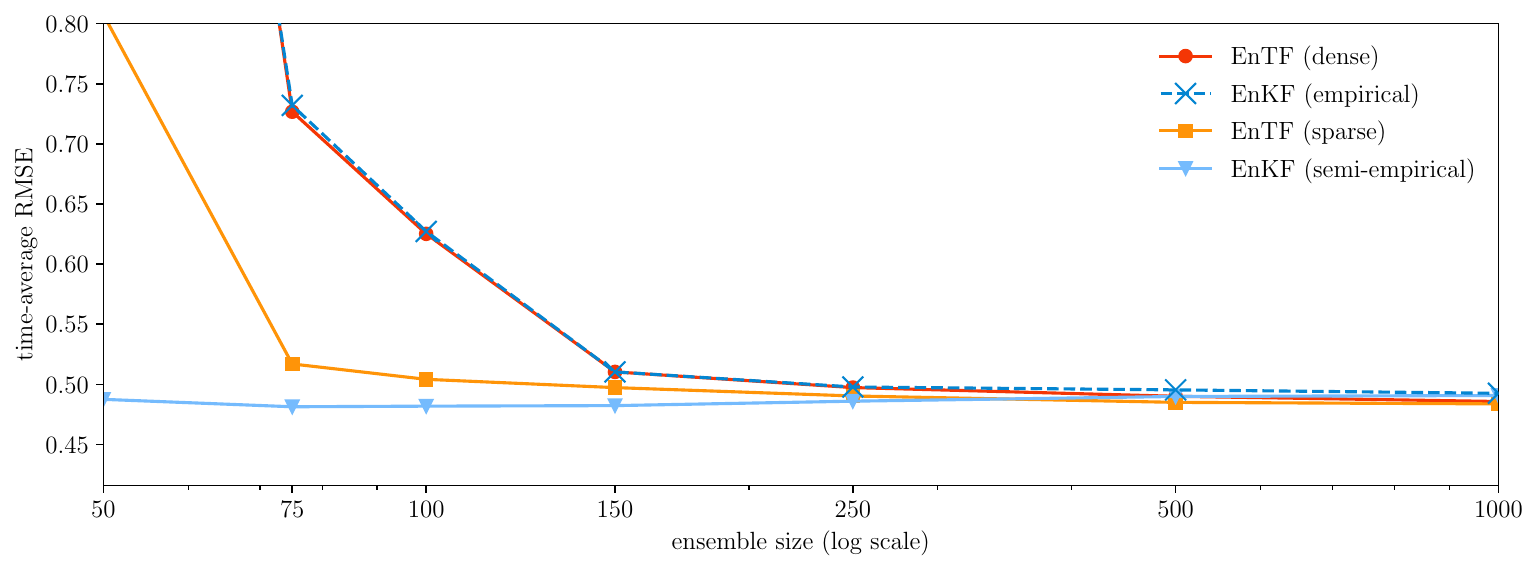}
  \caption{
 The time-average RMSE for different ensemble filters of the Lorenz-63 system. The sparse EnTF, which exploits conditional independencies using sequential updates and sparse low-dimensional maps, yields lower RMSE than an EnTF with dense maps and the equivalent stochastic EnKF, particularly in the small ensemble size regime.
  }
  \label{fig:filter_results}
\end{figure}

Figure~\ref{fig:filter_results} presents the average RMSE for the estimated mean of an ensemble transport filter (EnTF) that uses sparse sequential updates, as well as for two configurations of the EnKF: an empirical (fully sample-based) EnKF, and a semi-empirical EnKF (analogous to Equation~35). Following a spin-up period, the RMSE for estimating the true hidden state is evaluated over $1000$ time steps and the results are averaged over $100$ repeated simulations with different random seeds. 
We observe that explicitly accounting for conditional independence significantly improves filtering performance, especially for small ensemble sizes.  In particular, the ensemble transport filter based on sparse filter updates (EnTF-sparse) has lower time-averaged RMSE than using dense affine transport maps without accounting for sparsity. The performance of the latter filter (EnTF-dense) matches that of a stochastic EnKF, which similarly does not exploit conditional independence. Indeed, the stochastic EnKF constructs the composite map directly as a function of six-dimensional inputs. 
Furthermore, the sparse EnTF produces RMSEs that are close to those obtained with the semi-empirical EnKF for medium ensemble sizes, i.e. $N \geq 75$. Yet, the latter is given ``extra'' information: namely, the form of the observation operator and the covariance matrix of the observational noise. The transport filter and the stochastic EnKF must learn the update given only access to the ensemble of states and observations. 

\end{appendices}

\FloatBarrier

\selectlanguage{english}
\bibliographystyle{plainnat}
\bibliography{references}

\end{document}